\documentclass{article}
\usepackage[utf8]{inputenc}
\usepackage{style}

\title{Unbiased Estimation of Structured Prediction Error}
\date{}

\hypersetup{colorlinks=true,linkcolor=blue,urlcolor=blue}
\author{Kevin Fry}
\author{Jonathan E. Taylor}
\affil{\textit{Department of Statistics \\ Stanford University \\ Stanford, CA \\ email:} \href{mailto:kfry@stanford.edu}{\texttt{kfry@stanford.edu}}}
\begin{document}

\maketitle

\begin{abstract}
    Many modern datasets, such as those in ecology and geology, are composed of samples with spatial structure and dependence. With such data violating the usual independent and identically distributed (IID) assumption in machine learning and classical statistics, it is unclear \textit{a priori} how one should measure the performance and generalization of models. Several authors have empirically investigated cross-validation (CV) methods in this setting, reaching mixed conclusions. We provide a class of unbiased estimation methods for general quadratic errors, correlated Gaussian response, and \textit{arbitrary} prediction function $g$, for a noise-elevated version of the error. Our approach generalizes the coupled bootstrap (CB) of \citet{Oliveira2021} from the normal means problem to general normal data, allowing correlation both within and between the training and test sets. CB relies on creating bootstrap samples that are intelligently decoupled, in the sense of being statistically independent. Specifically, the key to CB lies in generating two independent ``views" of our data and using them as stand-ins for the usual independent training and test samples. Beginning with Mallows' $C_p$ \citep{Mallows1973}, we generalize the estimator to develop our \textit{generalized $C_p$} estimators (GC). We show at under only a moment condition on $g$, this noise-elevated error estimate converges smoothly to the noiseless error estimate. We show that when Stein's unbiased risk estimator (SURE) applies, GC converges to SURE as in the normal means problem. Further, we use these same tools to analyze CV and provide some theoretical analysis to help understand when CV will provide good estimates of error. Simulations align with our theoretical results, demonstrating the effectiveness of GC and illustrating the behavior of CV methods. Lastly, we apply our estimator to a model selection task on geothermal data in Nevada.

\end{abstract}


\section{Background}
    Prediction error estimation is of central importance to modern statistics, machine learning, and artifical intelligence. Understanding how well a model will predict on unseen data is crucial to the determination of whether it should be used, how it should be used, and to what extent humans can trust and rely on its predictions. In the standard setting, it is assumed that data points are IID or exchangeable, which is amenable to standard practices of data splitting and CV. However, in settings where the data is not IID, the answer becomes more complicated. Methods must account for the correlation between sample points that bias usual error estimates. However, in practice we see practitioners still employing CV for the purpose of prediction error estimation of spatial models \cite{Ploton2020, Wadoux2021}.
    
    \subsection{Problems with CV} \label{sec:cvprobs}
        We begin by reviewing CV methods. Recall that a CV method uses some procedure to split the dataset into $k$ folds, repeatedly trains the model on $k-1$ folds and gets an estimate of performance on the held out fold. The estimates on the held-out folds are averaged to produce the final estimate of performance. The most basic version of this is standard $k$-fold CV (KFCV), whereby observations are randomly assigned to each fold, only requiring each fold be the same size. However, this IID assumption breaks down in spatial settings. Aiming to remedy this, spatial $k$-fold CV (SPCV) splits the data into $k$ folds while keeping nearby points together, e.g. all observations from the same county are put in the same fold. The idea here is to cut down the correlation between folds by keeping spatially nearby points together. Going further with this idea, one can consider a buffered-leave-one-out CV (BLOOCV) approach, whereby each observation is its own fold, but we additionally remove observations within a spatial buffer of the validation observation from training.

        However, these spatial CV approaches cannot be blindly applied to spatial data problems. One may think this is due only to the varying correlation structure between differing training locations and testing locations. This is not the case. Even when the training and testing locations are the same CV has issues. An example of a such a situation would be measuring temperature at the same place at different times, while using time-invariant features for prediction. That is we have a $X,Y$ and $Y^*$ with samples indexed to the same locations. This sampling scheme, and others, are encapsulated by \cref{eq:models}. We demonstrate even here CV methods struggle with an ordinary least squares (OLS) example with $n=100$ and $p=5$. We generate $Y^*, Y$ according to the two  mechanisms in \cref{eq:models}. In the first, they are independent draws conditional on $X$ from $\cN(X\beta, \Sigma_Y)$, referred to as the no shared noise model (NSN). In the second, they are correlated conditional on $X$ and they are drawn according to \cref{eq:corrdatamodel} with non-zero $\Sigma_{Y,Y^*}$. This is referred to as the shared structural noise model (SSN). The error metric is the relative mean squared error, with (Monte-Carlo estimated) ground truth (red) as the baseline. Full simulation details are given in \cref{sec:sims}. 
        \begin{figure}[H]
            \centering
            \includegraphics[width=.95\linewidth]{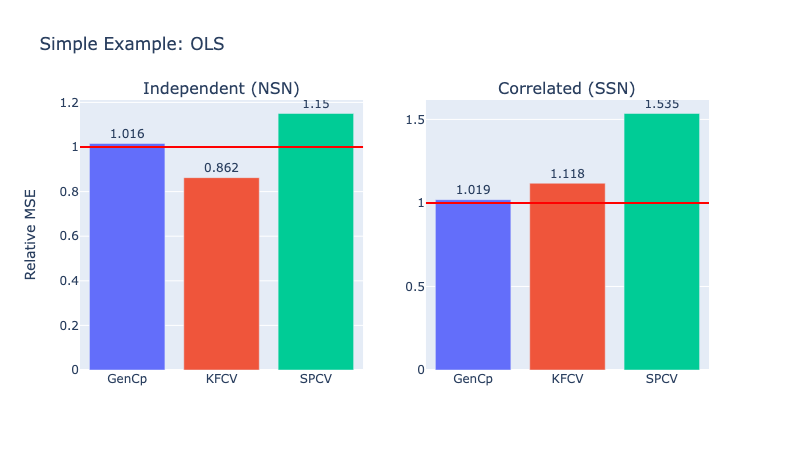}
            \caption{Relative MSE of adaptive linear regression for independent $Y^*, Y$. `GenCp' refers to our proposed method, specifically the variant given in \cref{eq:lincp}. `KFCV' is for standard $k$-fold CV and `SPCV' is spatial CV with $k$-means learned groups. Full details given in \cref{sec:sims}.}
            \label{fig:motiv_ex}
        \end{figure}

        It is clear both regular $k$-fold CV and spatial CV are biased estimators of the prediction error. We primarily focus on the NSN and SSN settings, but do discuss CV for general settings, including out-of-sample error estimation, in \cref{sec:whatcv}.

    \subsection{Setting and notation}
        We consider this problem in the setting of normally distributed data with arbitrary correlation where $X \in \R^{n \times p}$ is a common feature matrix that is considered fixed throughout and shared between $Y^*, Y \in \R^n$. Explicitly,
        \begin{equation} \label{eq:corrdatamodel}
            \begin{pmatrix}
            Y^* \\ Y
            \end{pmatrix} \mid X \sim \cN\left(\mu, \Sigma\right), \qquad \Sigma = \begin{pmatrix} \Sigma_{Y^*} & \Sigma_{Y^*, Y} \\ \Sigma_{Y,Y^*} & \Sigma_Y \end{pmatrix}
        \end{equation}
        And for a prediction function $g$ fit on $Y$ we want to estimate the generalized quadratic loss
        $$
        \E \|Y^* - g(Y) \|_{\Theta_p}^2 = \E[(Y^* - g(Y))^\top \Theta_p (Y^* - g(Y))]
        $$
        
        Under what setting would we arrive at such data? A useful class of data-generating mechanisms is the following. Consider a process $Y$, such that
        \begin{align*}
            Y_t(X_t) &= f(X_t) + \epsilon_t \\
            \epsilon_t &= \epsilon^S_t + \epsilon^M_t
        \end{align*}
        where $t \in \R^d$ is some spatial index, for instance $t$ could be time, locations on Earth, or regions of the brain. The superscripts $S$ and $M$ denote spatial noise (e.g. due to unmodeled features) and measurement noise, respectively. We then consider a replicate $Y^*$, which could take on either of the following forms
        \begin{align} \label{eq:models}
            Y^*_t(X_t) &= f(X_t) + \epsilon^*_t & \nonumber  \\
            \epsilon^*_t &= \epsilon^{S,*}_t + \epsilon^{M,*}_t  & \text{(NSN)} \\ 
            \epsilon^*_t &= \epsilon^{S}_t + \epsilon^{M,*}_t  & \text{(SSN)} \nonumber
        \end{align}
        Here we distinguish between the two models by referring to the former as the no shared noise (NSN) model, and the latter as the shared structural noise model (SSN) model. This is to emphasize NSN generates independent vectors $Y$ and $Y^*$ given $X$, while SSN generates vectors $Y, Y^*$ that share the structural component of their covariance matrices. Note here that we do not require $\epsilon_t, \epsilon^*_t$, or even their spatial and measurement components to be independent or identically distributed. Now for some sample of $n$ locations and model $g$, we want to estimate the prediction error, which for this paper we define as
        $$
        \mathrm{Err}(g) = \E \| Y^* - g(Y) \|_{\Theta_p}^2
        $$
        In words, we are interested in the MSE for a new replicate $Y^*$ with the same features $X$ at the same locations we have observations of $Y$ for. This can be viewed as the spatial analog to the in-sample error setting considered by Mallows' $C_p$ and other covariance penalty estimators \cite{Akaike1998, Efron1986, Efron2004, ESL, Mallows1973}. The generalization being that we allow $Y^*$ to be correlated with $Y$, which could arise due to unmodeled spatial variation, for instance.
        
    \subsection{Stein's unbiased risk estimator}
        One cannot write about unbiased risk estimation in normal data without mention of the seminal work of Charles Stein \citep{Stein1981}. The most well-known and widely applied version of the result provides an unbiased estimator of squared error for IID normal data $Y_i \sim \cN(\mu, \sigma^2)$, and weakly differentiable estimator $g$. This has been used to provide unbiased estimates of risk and prove the degrees of freedom of notable prediction functions, such as the LASSO in \citep{Tibshirani2012}. 
        A version of SURE exists for general Gaussian data, but unbiasedly estimates a generalized quadratic form matched to the covariance structure of the data. This whitens the data and in effect reduces it to the isotropic Gaussian case. 
        \begin{theorem} \label{thm:stein}
            Let $Y \sim \cN(\mu, \Sigma)$ with $\Theta = \Sigma^{-1}$. Let estimator $g : \R^n \mapsto \R$ be s.t. $f(z) = \Theta^{1/2} g(\Theta^{-1/2} z)$ is weakly differentiable where $\Theta^{1/2}$ is the symmetric square root. Write $\nabla \cdot g := \sum_i \nabla_{i} g_i$ for the divergence of $g$ with $\nabla_i g_j$ being the weak partial derivative of the $j$-th component of $g$ with respect to $y_i$. Assume that $\E\|g(Y)\|_\Theta^2 < \infty$ and $\E|\nabla_{i} g_i(Y)| < \infty$. Then,
            \begin{equation} \label{eq:corrsure}
                \mathrm{SURE}(g, \Theta) = \|Y - g(Y)\|_\Theta^2 + 2 (\nabla \cdot g)(Y)
            \end{equation}
            is an unbiased estimator of $\mathrm{Err}(g)$.
        \end{theorem}
        This is helpful in many cases but again requires weak differentiability and calculating a divergence term. Our proposed estimator does not make any assumptions on $g$ and works with any generalized quadratic $\|\cdot\|_Q^2$ for any $Q$. However, in \cref{sec:theory} we show that our estimator does converge to SURE when SURE is well defined.

        SURE has been extended to discontinuous functions by \citep{Tibshirani2015}. However, this extension relies on calculating the discontinuity sets of the function, which can be very difficulat. Despite this, \citet{Tibshirani2015} use a different approach to derives degrees of freedom for best subsets and relaxed lasso for IID normal response and orthogonal $X$. More recently, \citet{Oliveira2021} break from the SURE approach, proposing the coupled bootstrap (CB) to estimate error for general $g$ and IID response. 
    
    \subsection{Contributions}
        This paper has two main contributions. First, the most general form of generalized $C_p$ provides an unbiased estimate for generalized quadratic prediction error when the joint vector $(Y^*, Y)$ has arbitrary covariance structure $\Sigma$ and arbitrary prediction function $g$ fitted on noise-elevated data. We present several GC variants depending on the specific setting, but refer to them interchangeably as generalized $C_p$ or GC estimators since they fall into the same general framework and can be seen as generalizations of Mallows' $C_p$. Further, we prove the noise-elevated error smoothly converges to the noiseless error in \cref{prop:smooth}. Empirical results verify GC estimators with small added noise do indeed closely approximate the noiseless error. We also show in \cref{thm:gencptostein} that when SURE exists, GC converges to SURE. In addition, we develop the special case of estimation of prediction error of bagged arbitrary prediction function $g$. Second, we apply a similar analysis to CV methods to provide a theorem describing when CV would provide consistent error estimates. This theoretical result not only adds theoretical context explaining empirical results from \citep{Ploton2020,Wadoux2021}, but also provides guidelines for how practitioners should choose a CV method for their dataset and application.
    
        The structure of the rest of the paper is as follows:
        \begin{itemize}
            \item In \cref{sec:gencp} we develop GC methods for error (and therefore risk) estimation in general Gaussian data. We begin with adaptive linear smoothers, build into bagging these smoothers, and ultimately to general (bagged) models.
            We note that GC even allows for correlation between the observed $Y$ and unobserved $Y^*$, as well as $Y^*$ and $Y$ having different covariance structure. In \cref{sec:bag} we extend these results to bagged estimators, and discuss estimating the covariance $\Sigma$ in \cref{sec:estcov}.

            \item In \cref{sec:theory} we present theoretical results of GC estimators. We show that the noise-elevated error converges to the noiseless error as we reduce the added noise to 0. We also demonstrate GC recovers SURE int the limit when SURE exists. 
            
            \item In \cref{sec:optim} we extend the usual notion of optimism to our setting, and analyze the natural nonparamteric bootstrap estimators that arise from it, showing their insufficiency for the task.

            \item In \cref{sec:whatcv} we use our results to formally investigate what CV methods are estimating in the spatial setting, and how that differs from what one may think it is estimating. We use this insight to attempt to resolve the debate between \cite{Ploton2020} and \cite{Wadoux2021}, and give advice on how practitioners should choose a CV method for their particular application.

            
            \item In \cref{sec:sims} we present simulations validating the unbiasedness of our estimator and comparing it to CV methods. In \cref{sec:realdata} we use GC for model selection on geothermal data in Nevada. We also demonstrate that in practical settings, CV estimators deviate from GC.
            
            \item In \cref{sec:conclusion} we conclude, discussing future steps such as predicting local maxima, incorporating the approach in CV methods, and extending to other distributions.
        \end{itemize}

    \subsection{Related work}
        Estimating and understanding model performance is a classical problem in statistics and machine learning. In the above exposition, we have shown that CV struggles to accurately estimate error in the SSN setting as well as the easier NSN setting. Further we have established a notion of error that is an extension of in-sample error for the spatial setting. For IID data and in-sample error, covariance penalty approaches such as the classical Mallows' $C_p$ and AIC \citep{Akaike1998, Mallows1973}, with extensions to the ideas of optimism and degrees of freedom \citep{Efron1986, Efron2004, ESL} have given sufficient answers to estimating prediction error of a model. We frame our work as building off of Mallows', moving past the IID assumption to spatially correlated data.
        
        In the standard IID case, the foundational result of \citep{Mallows1973} offers a satisfactory answer for linear models, needing only an estimate of the noise level $\sigma$. We develop our approach with Mallows' $C_p$ as a starting point, extending it to the spatial settings of NSN and SSN defined in \cref{eq:models} for adaptive linear smoothers, which gives general predictive models as a corollary. Our approach can be seen as an elaboration on the use of the parametric bootstrap to estimate prediction error. There have been proposed methods using a parametric bootstrap to estimate risk in IID normal data \citep{Breiman1992, Efron2004, Ye1998}. However, these methods do not capture the full optimism, see \citet{Oliveira2021} for a succinct and clear discussion of this. We also show the spatial extenstions of these parametric bootstrap methods are also not fit-to-task in \cref{sec:optim}. \citet{Tian2020} estimates prediction error for linear models after feature selection. One can view this as a special case of \cref{thm:alincorrcp}, for IID Gaussian data and a subset of adaptive linear smoothers. Finally, our method addresses bagged estimators which previous works do not. \citet{Oliveira2021} generalizes \citet{Tian2020} to arbitrary model choice $g$, but again only in the IID setting. We discuss these works in more depth in \cref{sec:cbcomp}. Both \citep{Oliveira2021} and \citep{Tian2020}, as well as earlier works \citep{Robins2006,Tian2016}, leverage a trick for Gaussian vectors to generate two independent views with the same mean as the original vector. This has since then been named ``data fission" and generalized to other distributions by \citet{Leiner2021}, and the interested reader should refer to their paper for an extensive enumeration of ``fissions". In the present work, we leverage this popular Gaussian ``fission" trick as well. The authors of \citep{Oliveira2021} extend their results to the IID Poisson case, where an analogous trick applies, in \citep{Oliveira2022}. We give a more in-depth discussion of \citep{Oliveira2021,Tian2020} in \cref{sec:cbcomp}.

        Other authors have looked at other error metrics for CV methods and other covariance penalties  \citep{Bates2021,Rosset2020}. \citet{Bates2021} investigate CV in the IID setting, and find that the method does not estimate out-of-sample error, that is the expected error over a new draw $(x_0,y_0)$ for a model trained on the data at hand, but instead more closely estimates the average out-of-sample error over training sets $(X,Y)$. Although the majority of our paper is for the fixed-$X$ setting, in \cref{sec:whatcv} we derive insights for how CV performs both in-sample and out-of-sample in the spatial setting. These insights provide heuristics for how one should construct their CV procedure depending on the particulars of the dataset and problem. \citet{Rosset2020} takes a random-$X$ approach to in-sample error estimation, presenting a generalization to Mallows' $C_p$ they term $RC_p$. Although we also start with Mallows' $C_p$, our approach keeps to the fixed-$X$ regime and instead moves beyond the IID setting.

        As mentioned in the introduction, CV methods have been used by practitioners to estimate prediction error in spatial data.  \citet{Parmentier2011} propose the BLOOCV approach, where observations within a certain buffer radius of the held-out observation are not used in training. With a sufficiently large buffer, one would expect the training set to be independent of the held out observation, recovering the nice properties of the IID setting. However, if the future data points will be correlated with the training data, then BLOOCV will obviously overestimate error. As discussed above in \cref{sec:cvprobs}, CV approaches do not estimate prediction error well in the SSN setting. \citet{Ploton2020} applied CV methods, including BLOOCV, to spatial data, arguing spatial CV methods are necessary to not overestimate model performance. Soon after, \citet{Wadoux2021} responded to this, disagreeing and apparently demonstrating standard $k$-fold CV was a better estimate of performance than its spatial counterparts. Both of these papers are concerned with out-of-sample error. After presenting GC, we use the same tools to analyze CV estimators and derive a result explaining when CV should provide good estimates of out-of-sample error for correlated data in \cref{sec:whatcv}.

\section{Methodology} \label{sec:gencp}
    \subsection{Mallows' \texorpdfstring{$C_p$}{Cp}} \label{sec:mallows}
        Mallows' $C_p$ \cite{Mallows1973} is a classical result for getting an unbiased estimate of prediction error in linear regression, also known as ordinary least squares, for IID data. Formally, if
        $$(Y^*,Y) \mid X \overset{iid}{\sim} \cN(\mu, \sigma^2 I)$$
        and $P = X(X^\top X)^{-1}X^\top$ is the usual OLS estimate, then
        \begin{align*}
            \E \|Y^* - P Y\|_2^2
            &= \E \|Y - P Y\|_2^2 + 2p\sigma^2
        \end{align*}
        Although a useful result on its own, it falls short for our setting. It requires IID data, only gives you predictions for the same sites you use for fitting the model, and it does not hold for general models.
        
        It is easy enough to relax the IID assumption. If instead, $(Y^*, Y) \mid X \overset{ind}{\sim} \cN(\mu, \Sigma_Y)$, one can easily see the Mallows' unbiased estimator is then:
        $$
        \E \|Y^* - P Y\|_2^2 = \E \|Y - P Y\|_2^2 + \tr(2P \Sigma_Y)
        $$
        Now if we consider a general linear smoother $S$, we can arrive at an estimator of the same form:
        \begin{align*}
            \E \|Y^* - S Y\|_2^2 &= \E \|Y^* - Y \|_2^2 + \E \|Y - SY\|_2^2 + 2\E\left[(Y^* - Y)^\top(Y - SY)\right] \\
            &= 2\tr(\Sigma_Y) + \E \|Y - SY\|_2^2 - 2\E\left[(Y - \mu)^\top(I-S)(Y - \mu)\right] \\
            &= \E \|Y - SY\|_2^2 + 2\tr(S\Sigma_Y)
        \end{align*}
        
        Now we can add estimation and prediction matrices $\Theta_e, \Theta_p$ respectively. Here $\Theta_e$ is some augmentation matrix applied to the data before feeding it into the model. $\Theta_p$ is used to generalize our error matrix with a generalized quadratic $\|\cdot\|_{\Theta_p}$. A useful example is when $\Theta_e$ is the selector matrix for a training set $\cT$. For example, if $\cT = \{1,3\}$ and $n=4$, then
        $$
        \Theta_e = \begin{bmatrix} 1 & 0 & 0 & 0 \\ 0 & 0 & 1 & 0 \end{bmatrix}
        $$
        Then setting $\Theta_p = I - \Theta_e^\top \Theta_e$ we derive an estimator for a train/test split of our data:
        \begin{align} \label{eq:lincp}
            \E \|Y^* - S_e Y\|_{\Theta_p}^2 &= \E \|Y - S_e Y\|_{\Theta_p}^2 + 2\tr(\Theta_p S_e \Sigma_Y)
        \end{align}
        where $S_e$ denotes the smoother based on $\Theta_e X, \Theta_e Y$. For the rest of the paper, we often suppress the subscript $e$ notation as the results do not change for $S$ fit on the full training data or on the training subset.
        
        Above, we have constructed a Mallows' $C_p$-like estimator for structured data test set error for linear smoothing models. This is of some interest on its own, but still leaves out a vast range of models used by practitioners today.
    
    \subsection{Data ``fission'' by randomization}
        To generalize beyond non-adaptive linear smoothers, that is smoothers that are not dependent on $Y$, we need to introduce a simple but clever tool.
        
        A well known trick allows you to ``fission" a Gaussian vector $Y \sim \cN(\mu, \Sigma_Y)$ to generate two independent views with the same mean $\mu$. This approach has been used for various applications such as selective inference \citep{Tian2016} and nonparametric confidence sets \citep{Robins2006}. This method is expanded upon in \cite{Leiner2021}, from which we borrow the term data fission we will use throughout. By cleverly adding noise $\omega$:
        \begin{equation} \label{eq:rand}
            \begin{gathered}
                \omega \sim \cN(\mathbf{0}, \Sigma_Y), \qquad W = Y + \sqrt\alpha\omega \sim \cN(\mu, \Sigma_W), \qquad  W^\perp = Y - \omega / \sqrt\alpha \sim \cN(\mu, \Sigma_{W^\perp}) \\ 
                \Sigma_W = (1+\alpha)\Sigma_Y, \qquad \Sigma_{W^\perp} = (1+1/\alpha)\Sigma_Y
            \end{gathered}
        \end{equation}
        We note that one cannot recover $Y$ from either view alone, but with both we have the decomposition
        \begin{equation} \label{eq:randdecomp}
            \begin{gathered}
                Y = AW + A^\perp W^\perp \\
                A = I - A^\perp, A^\perp = (1+1/\alpha)^{-1} I
            \end{gathered}
        \end{equation}
        With this, we can derive a Mallows'-like unbiased estimate of $\mathrm{Err}(S) = \E \|Y^* - S(\Theta_e, W)Y \|_{\Theta_p}^2$ for a data-dependent linear smoother $S(\Theta_e, \cdot)$:
        \begin{theorem} \label{thm:alincp}
            With the above setup, an unbiased estimate of the prediction error $\E\|Y^* - S(W)Y \|_{\Theta_p}^2$ is given by
            \begin{equation} \label{eq:traceestrfull}
                \|W^\perp - S(W)Y \|_{\Theta_p}^2 + \tr(\Theta_p\Sigma_Y) - \tr(\Theta_p\Sigma_{W^\perp}) + 2\tr(\Theta_p S(W)\Sigma_Y)
            \end{equation}
            and an unbiased estimate of $\E\|Y^* - S(W)W \|_{\Theta_p}^2$ is given by
            \begin{equation} \label{eq:traceestr}
                \|W^\perp - S(W)W \|_{\Theta_p}^2 + \tr(\Theta_p\Sigma_Y) - \tr(\Theta_p\Sigma_{W^\perp})
            \end{equation}
        \end{theorem}
        The proof is an exercise in linear algebra and deferred to \cref{sec:alinproof}. This has a nice interpretation. The first two terms are adjustments for the differences in noise level between $Y^*$ and $W^\perp$, and the last term is the usual Mallows' correction, now with $S(W)$ random. We present refitting on $W$ and $Y$ as precursors to our bagging results \cref{thm:bagging,thm:corrbagging}, where $S(W)W$ is the conventional refit for each learner in the bag, and $S(W)Y$ is an alternative that intuitively should provide better predictions.

        We note that \cref{thm:alincp} we need not use the trace correction, but any quantity with that expectation. A natural choice is to extend the choice of \citet{Oliveira2021} to our setting, which they showed has smaller variance than other options:
        \begin{equation}\label{eq:randestrfull}
            \mathrm{GC}^{\indep} (Y) := \|W^\perp - S(W)Y \|_{\Theta_p}^2 + \tr(\Theta_p\Sigma_Y) - \tr(\Theta_p\Sigma_{Y^*}) - \|\omega\|_{\Theta_p}^2 / \alpha + 2\tr(\Theta_p S(W)\Sigma_Y)
        \end{equation}
        \begin{equation}\label{eq:randestr}
            \mathrm{GC}^{\indep} (W) := \|W^\perp - S(W)W \|_{\Theta_p}^2 + \tr(\Theta_p\Sigma_Y) - \tr(\Theta_p\Sigma_{Y^*}) - \|\omega\|_{\Theta_p}^2 / \alpha
        \end{equation}
        We verify this random correction has smaller variance than the trace correction in \cref{fig:rand_vs_trace}.
        
        The above is for one realization of $W, W^\perp$. We can repeat this for many such bootstrap samples $W_b, W_b^\perp$ and average them to get a better estimate. This is what has been called the coupled bootstrap (CB) by \citet{Oliveira2021}. Despite the name, this technique in fact cleverly creates marginally decoupled samples, in the sense that they are statistically independent. So we have the following corollary 
        \begin{corollary},
            In the setting of \cref{thm:alincp}
           $$\frac{1}{B} \sum_{b=1}^B \mathrm{GC}_b^{\indep}(W_b)$$
            is an unbiased estimate of $\E \|Y^* - S(W)W\|_{\Theta_p}^2$ and 
            $$\frac{1}{B} \sum_{b=1}^B \mathrm{GC}_b^{\indep}(Y)$$
            is an unbiased estimate of $\E \|Y^* - S(W)Y\|_{\Theta_p}^2$.
        \end{corollary}
        
        Thus we suggest generating $W, W^\perp$ for small $\alpha$, and using the above bootstrapped estimator of the error of $S(W)W$ to get an estimate of the estimators $S(Y)Y$. In all our applications and simulations, we employ this bootstrap version of the estimator. 

        We claim the above applies to decision trees. To see a decision tree is an adaptive linear smoother is in fact quite simple. Conditional on the data it was fit on, a tree is simply a function $S(\cdot)$ that groups observations based on their features, and predicts the mean of the observations in each group. That is,
        $$
        S(W) = \begin{pmatrix}
            n_1^{-1}\1_{n_1} \1_{n_1}^\top & 0 & \cdots & \cdots & 0 \\
        0 & \ddots & \ddots & & \vdots \\ \vdots & \ddots & n_\ell^{-1}\1_{n_\ell}\1_{n_\ell}^\top & \ddots & \vdots \\
        \vdots & & \ddots & \ddots & 0 \\
        0 & \cdots & \cdots & 0 & n_L^{-1} \1_{n_L}\1_{n_L}^\top
        \end{pmatrix}
        $$

    \subsection{Correlated \texorpdfstring{$Y, Y^*$}{Y, Y*}} \label{sec:corrtraintest}
        Our first main result \cref{thm:alincorrcp} moves beyond the case of $Y^*$ and $Y$ being IID vectors conditional on $X$ and simply considers them to be a jointly Gaussian vector.
            
        It uses the fact that we can regress normal vector $Y^*$ on $Y$ to obtain the decomposition:
        $$
        Y^* = N + \Gamma Y, \qquad \Gamma = \Sigma_{Y^*, Y}\Sigma_{Y}^{-1},\qquad N = Y^* - \Gamma Y,\qquad \Delta = (S(W) - \Gamma) Y, \qquad N^\perp = (I - \Gamma) W^\perp
        $$
        Note that $N \indep Y$. Similarly define 
        $$
        \Gamma_W = \Sigma_{Y^*, W}\Sigma_{W}^{-1}, \qquad N^* = Y^* - \Gamma_W W, \qquad N^W = (I - \Gamma_W) W^\perp, \qquad \Delta^W = (S(W) - \Gamma_W) W
        $$
        \begin{theorem} \label{thm:alincorrcp}
            Suppose 
            \begin{equation*}
                \begin{pmatrix} Y^* \\ Y \end{pmatrix} \mid X \sim \cN\left(\mu, \Sigma \right), \qquad \Sigma = \begin{pmatrix} \Sigma_{Y^*} & \Sigma_{Y^*, Y} \\ \Sigma_{Y, Y^*} & \Sigma_Y \end{pmatrix}
            \end{equation*}
            and denote randomized views $W, W^\perp$ of $Y$ defined as in \cref{eq:rand}. Then,
            \begin{equation} \label{eq:correstrfull}
                \mathrm{GC}^{cor}(Y) := \|N^\perp - \Delta \|_{\Theta_p}^2 + \tr(\Theta_p (\Sigma_N - \Sigma_{(I - \Gamma)Y})) - \|(I - \Gamma)\omega\|_{\Theta_p}^2 / \alpha + 2\tr\left((I - \Gamma)^\top \Theta_p (S(W)-\Gamma) \Sigma_Y \right)
            \end{equation}
            is an unbiased estimate of $\E\|Y^* - S(W)Y \|_{\Theta_p}^2$, and
            \begin{equation} \label{eq:correstr}
                \mathrm{GC}^{cor}(W) := \|N^W - \Delta^W \|_{\Theta_p}^2 + \tr(\Theta_p (\Sigma_{N^*} - \Sigma_{(I - \Gamma_W)Y})) - \|(I - \Gamma_W)\omega\|_{\Theta_p}^2 / \alpha
            \end{equation}
            is an unbiased estimate of $\E\|Y^* - S(W)W \|_{\Theta_p}^2$.
        \end{theorem}

        The proof can be found in \cref{sec:alincorrproof}. Essentially we regress out the correlations between $Y^*$ and $Y$, and then apply the same approach as in \cref{thm:alincp}. And again we immediately have a bootstrapped version of the estimator:
        \begin{corollary}
            In the setting of \cref{thm:alincorrcp}
            $$\frac{1}{B} \sum_{b=1}^B \mathrm{GC}_b^{cor}(W_b)$$
            is an unbiased estimate of $\E \|Y^* - S(W)W\|_{\Theta_p}^2$ and 
            $$\frac{1}{B} \sum_{b=1}^B \mathrm{GC}_b^{cor}(Y)$$
            is an unbiased estimate of $\E \|Y^* - S(W)Y\|_{\Theta_p}^2$.
        \end{corollary}

    \subsection{Bagged estimators} \label{sec:bag}
        Bootstrap aggregating (bagging) is a common approach to create an ensemble model from training many base models on different bootstrap samples of the data, originally proposed by \citet{Breiman1996}. Bagging as been noted for its ability to smooth out discontinuous estimators, hence it also being known as bootstrap smoothing \cite{Buja2016,Efron2014}.
        \subsubsection{Uncorrelated \texorpdfstring{$Y, Y^*$}{Y, Y*}}
            In the special case where $\Sigma_\omega = \Sigma_Y$, one may view the fissioned view $W$ of our data as a parametric bootstrap with maximally-flexible or minimally-biased model as our mean, which is simply $Y$ itself. Thus it is natural to consider fitting $K$ models to fissioned views $W_k$ and estimating the error of the resulting bagged estimator:
            $$
            \E\left[\left\|Y^* - \frac{1}{K}\sum_{k=1}^K S(W_k)W_k \right\|_{\Theta_p}^2 \ \middle\vert\  X\right],
            $$
            or a modified bagged estimator that does a ``full refit":
            $$
            \E\left[\left\|Y^* - \frac{1}{K}\sum_{k=1}^K S(W_k)Y \right\|_{\Theta_p}^2 \ \middle\vert\ X\right],
            $$
            It is in fact quite straight-forward to extend \cref{thm:alincp} to the bagged setting. When viewed in the appropriate way, it is a simple linear algebra exercise deferred to \cref{sec:baggingproof}.
            \begin{theorem} \label{thm:bagging}
                Let $W_k, W_k^\perp$ be fissioned views of $Y$ from independent $\omega_k$'s, and consider the bagged estimator $K^{-1} \sum_k S(W_k)Y$. Then an unbiased estimate of $\E\left\|Y^* - \frac{1}{K}\sum_{k=1}^K S(W_k)Y \right\|_{\Theta_p}^2$ is
                $$
                K^{-1} \left[\sum_{k=1}^K \mathrm{GC}^{\indep}_k(Y) - \tr\left( (Z_K^{\perp})^\top \Theta_p Z_K^\perp\right) \right]
                $$
                and an unbiased estimate of $\E\left\|Y^* - \frac{1}{K}\sum_{k=1}^K S(W_k)W_k \right\|_{\Theta_p}^2$ is
                $$
                K^{-1} \left[\sum_{k=1}^K \mathrm{GC}^{\indep}_k(W_k) - \tr\left( (Z_K^{\perp})^\top \Theta_p Z_K^\perp\right) \right]
                $$
                where the subscript $k$ in $\mathrm{GC}^{\indep}_k$ denotes using the $k$-th fissioned view $W_k, W_k^\perp$ and $ Z_K^\perp = Z_K(I - K^{-1}\1_K\1_K^\top)$ is the vector of centered predictions.
            \end{theorem}
        \subsubsection{General \texorpdfstring{$Y,Y^*$}{Y,Y*}}
            Just as in the case of traditional estimators, we also have a correction for bagged estimators in the case where $Y$ and $Y^*$ are correlated. We simply rewrite
            $$
            \E\left\|Y^* - \frac{1}{K}\sum_{k=1}^K S(W_k)Y \right\|_{\Theta_p}^2 = \E\left\|N - \frac{1}{K}\sum_{k=1}^K\Delta_k \right\|_{\Theta_p}^2
            $$
            where $\Delta_k = (S(W_k) - \Gamma)Y$. Then the proof from \cref{thm:bagging} applies using the estimator from \cref{thm:alincorrcp}:
            \begin{theorem}\label{thm:alincorrbagging}
                Assume the setting of $Y \not \indep Y^* \mid X$ as in \cref{sec:corrtraintest}. Let $W_k, W_k^\perp$ be fissioned views of $Y$ from independent $\omega_k$'s, and consider the bagged estimator $K^{-1} \sum_k S(W_k)Y$. Then an unbiased estimate of $\E \left\| Y^* - \frac{1}{K} \sum_{i=1}^K S(W_k)Y \right\|_{\Theta_p}^2$
                is
                $$
                K^{-1} \left[\sum_{k=1}^K \mathrm{GC}^{cor}_k(Y) - \tr\left( (Z_K^{\perp})^\top \Theta_p Z_K^\perp\right) \right]
                $$
                and an unbiased estimate of $ \E \left\| Y^* - \frac{1}{K} \sum_{i=1}^K S(W_k)W_k \right\|_{\Theta_p}^2$
                is
                $$
                K^{-1} \left[\sum_{k=1}^K \mathrm{GC}^{cor}_k(W_k) - \tr\left( (Z_K^{\perp})^\top \Theta_p Z_K^\perp\right) \right]
                $$
                where the subscript $k$ in $\mathrm{GC}^{cor}_k$ denotes using the $k$-th fissioned view $W_k, W_k^\perp$ and $ Z_K^\perp = Z_K(I - K^{-1}\1_K\1_K^\top)$ is the vector of centered predictions.
            \end{theorem}
            The proof is identical to that of \cref{thm:bagging} and so we will not repeat it. Readers will notice this is all for parametric bagging, as opposed to the more often used nonparametric bagging. One may wonder if these parametric bagged models will perform similarly to the nonparametric bagged models. We investigate this in the case of random forests (RF) in \cref{sec:rfcomps}. In our simulations, we show that they do indeed perform similarly in terms of expected prediction error. 
            
            It is also worth noting that there is not a version of the nonparametric bootstrap developed for our setting, and so a parametric bootstrap is also more fit-to-purpose. There are several existing works proposing modifications to RF to make it better suited to spatial data \citep{Fox2020,Hengl2018,Saha2020}. \citet{Hengl2018} adds distance features to account for spatial correlation. \citet{Fox2020} sums an RF prediction and a regression kriging prediction on the residuals of RF. \citet{Saha2020} assumes an autoregressive structure on the covariance matrix to allow GLS-style fitting of trees in RF. It would be interesting in future to see how our parametric RF compares to these approaches.
            

    \subsection{General models}
        Our initial focus on linear adaptive smoothers is interesting for what it allows us to do with random forests as discussed above, which is interesting on its own. However, if one looks carefully at our proofs, you will see that when we refit on $W$, we need not use the fact that our model is a linear adaptive smoother. One can easily see this by noting that for any model $g$ we can define $S(W) = \mathrm{diag}(g(W))\mathrm{diag}(W)^{-1}$, and thus $S(W)W \eas g(W)$. Therefore we can provide estimators under \textit{both arbitrary $g$ and $Y \not\indep Y^* \mid X$.}
        \begin{corollary} \label{cor:corrgencp}
            Assume the setting of $Y \not \indep Y^* \mid X$ as in \cref{sec:corrtraintest}. Additionally let $\Delta_g = g(W) - \Gamma_W W$. Then an unbiased estimate of $\mathrm{Err}(g)$ is given by
            $$
            \mathrm{GC}^g(W) = \E \|N^\perp - \Delta_g \|_{\Theta_p}^2 + \tr(\Theta_p (\Sigma_{N^*} - \Sigma_{(I - \Gamma_W)Y})) - \|(I - \Gamma_W)\omega\|_{\Theta_p}^2 / \alpha
            $$
        \end{corollary}
        And so bootstrapping this also provides an unbiased estimator:
        \begin{corollary} \label{cor:corrcb}
            In the setting of \cref{cor:corrgencp}
            $$\frac{1}{B} \sum_{b=1}^B \mathrm{GC}_b^{g}(W)$$
            is an unbiased estimate of $\mathrm{Err}(g)$.
        \end{corollary}
        \cref{cor:corrcb} then gives us a direct generalization of the coupled bootstrap approach of \citep{Oliveira2021} to the SSN model with arbitrarily correlated Gaussian data.
        
        Of course \cref{cor:corrgencp} also then means we can immediately extend our result to \textit{bagging general models $g$ when $Y \not\indep Y^* \mid X$}. Now letting
        $$
        \mathbb{R}^{n \times K} \ni Z_K = 
        \begin{pmatrix} \vert & \vert & \vert \\
        \cdots & Y^* - g(W_k) & \cdots \\ \vert & \vert & \vert \end{pmatrix},
        \qquad \R^n \ni \bar Z_K = Y^* - \frac{1}{K} \sum_{k=1}^K g(W_k)
        $$
        we have the following theorem.
        \begin{theorem}\label{thm:corrbagging}
                Assume the setting of $Y \not \indep Y^* \mid X$ as in \cref{sec:corrtraintest}. If $W_k, W_k^\perp$ are fissioned views of $Y$ from independent $\omega_k$'s, and consider the bagged estimator $K^{-1} \sum_k g(W_k)$. Then an unbiased estimate of $ \E \left\| Y^* - \frac{1}{K} \sum_{i=1}^K g(W_k) \right\|_{\Theta_p}^2$
                is
                $$
                K^{-1} \left[\sum_{k=1}^K \mathrm{GC}^g(W_k) - \tr\left( (Z_K^{\perp})^\top \Theta_p Z_K^\perp\right) \right]
                $$
                where $ Z_K^\perp = Z_K(I - K^{-1}\1_K\1_K^\top)$ is the vector of centered predictions.
            \end{theorem}

\begin{figure} [H]
            \centering
            \includegraphics[width=.95\linewidth]{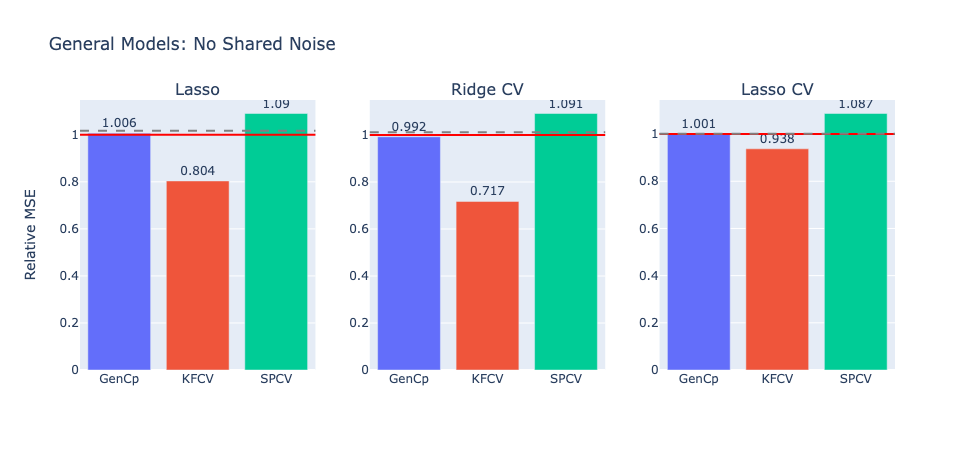}
            \caption{Relative MSE of general prediction functions for the NSN model.}
            \label{fig:gen_ind}
        \end{figure}

\begin{figure} [H]
            \centering
            \includegraphics[width=.95\linewidth]{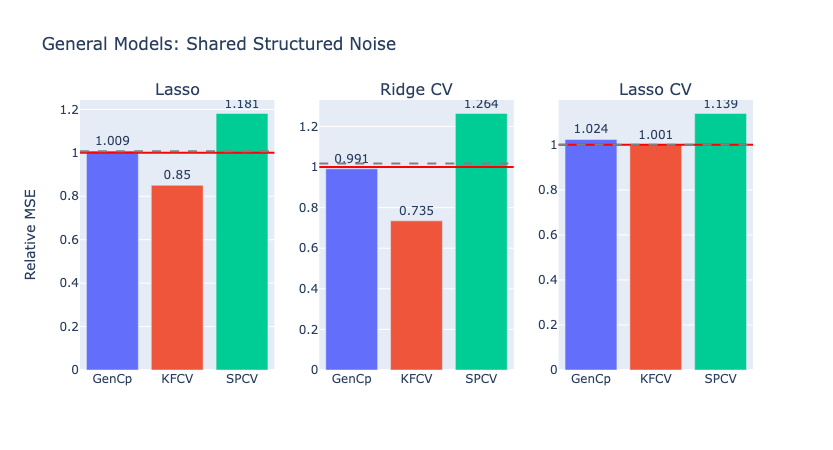}
            \caption{Relative MSE of general prediction functions for the SSN model.}
            \label{fig:gen_corr}
        \end{figure}

        This lets our estimator extend prediction functions such as the lasso or even CV-tuned lasso and ridge. We simulate these estimators for independent and correlated data vectors in \cref{fig:gen_ind} and \cref{fig:gen_corr} respectively, achieving almost unbiased results.

    \subsection{Estimation of covariance matrix} \label{sec:estcov}
        In practice we will often not know $\Sigma$ \textit{a priori} and thus must estimate it. The standard method for this problem is to assume the covariance function $C(s,t)$ belongs to some parametrized model, and fit those parameters to the residuals after fitting some prediction function. Typically such models depend only on the distance between locations $C(d)$ for $d=\|s - t\|_2$. For example, a widely used and effective model is the Matern covariance model. It is defined as
        \begin{equation}
            C(d) = b + \sigma^2 \frac{2^{1-\nu}}{\Gamma(\nu)} \left( \sqrt{2\nu} \frac{d}{\rho}\right)^\nu K_\nu \left(\sqrt{2\nu}\frac{d}{\rho}\right)
        \end{equation}
        where $b, \nu, \rho, \sigma^2$ are the parameters of the Matern model, $\Gamma(\cdot)$ is the Gamma function, and $K_\nu(\cdot)$ is the modified Bessel function of the second kind. An in-depth discussion of the Matern model is given in \citet{Minasny2005}. One notable strength of this model is depending on parameter values, one can go from an Exponential covariance function to a Gaussian covariance function. Software packages exist for fitting a Matern model, such as the Python package \texttt{scikitgstat} described in \citep{Malicke2021}, which is the package we use in this paper. They fit the Matern model with a bounded parameter version of the Levenberg-Marquadt algorithm \citep{Levenberg1944, Marquardt1963}.

        Once we have fitted a Matern model, we can use it and the locations $\{\ell_i\}$ of our data $Y$ to construct our covariance matrix $\Sigma$. We simply set $\Sigma_{ij} = C(\|\ell_i - \ell_j\|_2)$. If we are considering SSN, then we need to separate out the measurement noise from the structured noise. Fortunately, the $b$ in the Matern model is exactly the measurement noise, known as the \textit{nugget} in spatial statistics. We simply subtract it out from the diagonal to get our structural noise covariance matrix.

        An important decision is which prediction function to fit to the data to generate residuals in the first place. We do not offer theoretical analysis of this choice here but have found an approach that works well empirically. We find a good choice is to use the same prediction function to generate residuals for covariance estimation that one aims to estimate prediction error for.
        
\begin{figure} [H]
            \centering
            \includegraphics[width=.95\linewidth]{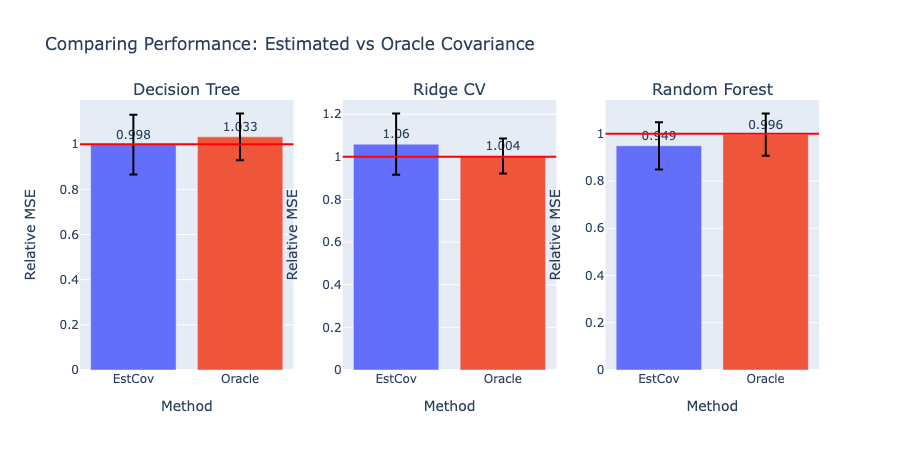}
            \caption{Comparing relative MSE using an estimated covariance matrix and an oracle. We plot results for a depth 3 decision tree as well as ridge tuned by cross-validation and random forest. EstCov (blue) refers to the estimates generated using a covariance matrix estimated from the data, while Oracle (red) refers to estimates generated with the ground-truth covariance matrix.}
            \label{fig:est_cov}
        \end{figure}

        Obviously we pay some price in doing this estimation, however \cref{fig:est_cov} show the price paid is not too large, at most 5\%. We simulated independent data vectors and used decision trees, ridge regression tuned by CV (RidgeCV), and RF as our models.

    \subsection{Relationship to Tian and Oliveira} \label{sec:cbcomp}
        The present work is quite related to \citet{Oliveira2021} and our estimators can be seen as generalizing CB to arbitrarily distributed normal response with common mean. \citet{Oliveira2021} considered the IID normal case for almost arbitrary prediction function $g$, building upon the earlier work of \citet{Tian2020} which only considered linear models with feature selection of the form $S(W)Y$. Our most general result is \cref{cor:corrcb}, which provides a direct generalization of \citet{Oliveira2021} to general Gaussian data. We also present \cref{thm:alincp,,thm:alincorrcp} which extend \citet{Tian2020} to general Gaussian data on our way to our bagging result \cref{thm:bagging,,thm:alincorrbagging,,thm:corrbagging}, and its utility compared to the IID CB presented in \citet{Oliveira2021} is worth discussing. 
        
        When considering a general model $g$, it is statistically better to choose $\alpha$ small, minimizing the perturbation to the model fit, and then shrinking the resulting large perturbations in $W^\perp$ through bootstrapping. However, this is approach is suboptimal when $g$ is a bagged model. By taking the approach of fitting the base estimator in a bagged model on the bootstrap samples $W_k$ with a sizeable $\alpha$, we can greatly increase computational efficiency, even in the IID case. Note that \cref{thm:bagging,,thm:alincorrbagging} only require a single round of bootstrapping. If we applied the standard CB estimator of \citet{Oliveira2021} to a bagged model, we would bootstrap once in computing CB estimator, and again in the fitting of the bagged estimator $g$. Thus our results offer a way to generate unbiased estimates for bagged models not only under a more general data generating process, but also with a significant computational speedup (proportional to the number of bootstrap samples).
        
        \citet{Oliveira2021} do include a small discussion about general covariance $\Sigma$ and generalized quadratics $\|\cdot\|_A$. However, this is still for the case where $Y^*, Y$ are IID conditional on $X$, i.e. NSN in \cref{eq:models}. Our work explicitly allows for correlation between $Y^*, Y$ via the SSN model thanks to \cref{thm:alincorrcp,,cor:corrgencp}. This is crucial for spatial settings, as it is unreasonable to assume the test points $Y^*$ are independent of the training points $Y$.

\section{Theoretical properties} \label{sec:theory}
    Our theoretical results demonstrate our method unbiasedly estimates prediction error for models fit on noise-elevated data $W$. Simulations show that choosing $\alpha$ close to 0 results in estimates close to that fit on the original data $Y$. Here we provide some theoretical results backing up these empirically observed properties. These are all straight-forward extensions of results from \citet{Oliveira2021} to the non-IID setting.

    First we can show the noise-elevated expected error
    $$\mathrm{Err}_\alpha(g) = \E \|Y^* - g(W_\alpha)\|_2^2$$
    is smooth in $\alpha$ as $\alpha \to 0$, where $W_\alpha = Y + \sqrt \alpha \omega$. In this section only, we use this subscript notation to denote the scale of the noise elevation as it becomes necessary to distinguish between different amounts of noise elevation.
    \begin{proposition}\label{prop:smooth}
        If for some $\beta > 0$ and integer $k \ge 0$,
        $$\E\left[\|g(W_\beta)\|_{\Theta_p}^2 \|W_\beta - \mu\|_{\Sigma_{Y}^{-1}}^{2m} \right] < \infty, \qquad m=0,\dots, k$$
        then $\alpha \mapsto \mathrm{Err}_\alpha (g)$ has $k$ continuous derivatives on $[0,\beta)$.
    \end{proposition}
    The proof is deferred to \cref{sec:smoothproof}. We repeat an important point made in \citet{Oliveira2021}, which is that the above result is achieved without assuming continuity of $g$. Intuitively, the additional Gaussian noise smooths out the discontinuities in $g$. Further, in our version of the theorem one may work with any generalized quadratic $\Theta_p$ for the measure of error.

    We can also analyze what happens to the estimator in the case where we have infinite bootstrap samples and take $\alpha \to 0$, that is look at $\lim_{\alpha \to 0} \mathrm{GCP}_\alpha^\infty (g)$ for
    $$\mathrm{GCP}_\alpha^\infty(g) = \E\left[\|W_\alpha^\perp - g(W_\alpha)\|_{\Theta_p}^2 - \|\omega\|_{\Theta_p}^2 / \alpha \mid Y\right]$$
    In \citet{Oliveira2021} they show that when SURE exists, the above quantity converges to SURE as $\alpha \to 0$. As one would hope, this results extends to our setting as well. The proof is a simple extension of the theorem in the IID case, thus we do not provide a proof here but interested readers should refer to \citet{Oliveira2021}.
    \begin{theorem} \label{thm:gencptostein}
          Consider the setting of \cref{thm:stein}, but at the $\beta$ noise-elevated level. That is $\E\|g(W_\beta)\|_\Theta^2 < \infty$ and $\E|\nabla_i f_i(\Theta^{1/2}Y)| < \infty$. Then
          $$\lim_{\alpha \to 0} \mathrm{GCP}_\alpha^\infty(g) \eas \mathrm{SURE}(g, \Theta)$$
          Thus the limit of $\mathrm{GCP}_\alpha^\infty(g)$ is unbiased for $\mathrm{Err}(g)$.
    \end{theorem}
    In summary, our proposed estimators provide unbiased estimates of the noise-elevated error $\mathrm{Err}_\alpha$, which converges to target error $\mathrm{Err}$ as $\alpha\to 0$. Furthermore, when SURE exists, our limiting estimator agrees with SURE.

\section{Optimism and parametric bootstrap estimators} \label{sec:optim}
    Although we present our results as stemming from Mallows' $C_p$, we take time here to relate them to more general notions in covariance penalty approaches. Recall Efron's optimism theorem \citep{Efron1986,Efron2004} in the IID setting:
    \begin{equation} \label{eq:optim}
        \E \|Y^* - g(Y) \|_2^2 = \E\|Y - g(Y)\|_2^2 + 2\sum_{i=1}^n \cov(Y_i, g(Y_i)) = \E\|Y - g(Y)\|_2^2 + 2\tr(\cov(Y, g(Y)))
    \end{equation}
    Estimating this covariance term is how approaches such as Mallows' and AIC work. In the spatial setting where $Y^*, Y \overset{ind}{\sim} P$, the decomposition remains the same, but in the more general setting where $Y^*, Y$ are correlated with eachother and not necessarily IID vectors, the error decomposees as:
    \begin{equation} \label{eq:corrspatoptim}
        \E \|Y^* - g(Y) \|_2^2 = \E\|Y - g(Y)\|_2^2 + 2\tr\left(\cov(Y, g(Y))\right) - 2\tr\left(\cov(Y^*, g(Y))\right) + \tr\Sigma_{Y^*} - \tr\Sigma_Y
    \end{equation}
    
    In light of the above, the approach of this paper can be seen as estimating this generalized optimism term for Gaussian data. Importantly the optimism is no longer a function of only the training data, but also of the new realization $Y^*$ through its covariance with the predictions.

    With this we can now examine the estimators for IID data in the literature \citep{Breiman1992,Efron2004,Ye1998} and extend them to our setting. A clear and more detailed explanation for the IID case is given in \citet{Oliveira2021}, but we present a brief summary here borrowing similar notation. These works estimate the optimism in \cref{eq:optim} with a parametric bootstrap. For a parametric bootstrap sample $ Y^{(b)} \mid Y \sim \cN(h(Y), \alpha\sigma^2 I_n)$ for some initial estimate $h(Y)$ and $\alpha \in (0,1]$. The works cited above choose $h(Y) = Y$. Then using the bootstrap estimate of covariance:
    \begin{equation*}
        \widehat{\cov} = \tr\left((B-1)^{-1} \sum_{b=1}^B g(Y^{(b)})(Y^{(b)} - \bar Y^{(B)})^\top\right)
    \end{equation*}
    we can construct the Efron (Efr) and Breiman-Ye (BY) estimates of error:
    \begin{equation*}
        \mathrm{Efr}_\alpha(g) = \|Y - g(Y)\|_2^2 + 2\widehat{\cov}, \qquad \mathrm{BY}_\alpha(g) = \|Y - g(Y)\|_2^2 + \frac{2}{\alpha}\widehat{\cov}
    \end{equation*}
    We note that \citet{Efron2004} specifically proposes $\mathrm{Efr}_\alpha$ with $\alpha=1$, but we present it in this more general form. With $\alpha=1$ we have $\mathrm{Efr}_\alpha = \mathrm{BY}_\alpha$. Now $\widehat{\cov}$ is intuitively an unbiased estimate of $\cov(Y^{(b)}, g(Y^{(b)}))$, and is argued for small $\alpha$ to also be close to $\cov(Y, g(Y))$. The essential shortcoming in the IID case is that $\widehat{\cov}$ is computed conditional on $Y$, and hence only captures part of $\cov(Y^{(b)}, g(Y^{(b)}))$, in particular:
    $$\E[\widehat{\cov} \mid Y] = \tr\left(\E\left[\cov(Y^{(b)}, g(Y^{(b)}) \mid Y) \right]\right)$$
    and thus $\E[\widehat{\cov}]$ is just the expected conditional covariance, while by the law of total covariance:
    $$\cov(Y^{(b)}, g(Y^{(b)})) = \E[\cov(Y^{(b)}, g(Y^{(b)}) \mid Y)] + \cov(\E[Y^{(b)} \mid Y], \E[g(Y^{(b)}) \mid Y] )$$
    Thus it is clear $\mathrm{Efr}_\alpha$ will be biased for noise-elevated covariance for small $\alpha$. The behavior of $\mathrm{BY}_\alpha$ is nice in special cases such as linear smoothers, but in general it is not clear what it is estimating or how to pick $\alpha$. The same results hold moving to a setting where $Y^* \indep Y$ but $Y \sim \cN(\mu, \Sigma_Y)$. Therefore, these estimators do not estimate the noise-elevated error, and it is not clear what they do estimate.

    Now if we move to the SSN model where $Y^*$ and $Y$ are dependent, we need to use the parametric bootstrap to instead estimate \cref{eq:corrspatoptim}. So we have to additionally estimate $\cov(Y^*, g(Y))$. A straightforward approach would be to use two sets of bootstrap samples $Y^{(b)}, Y^{*(b)}$ such that
    \begin{equation*}
        \begin{pmatrix}
            Y^{*(b)} \\ Y^{(b)}
        \end{pmatrix}
        \mid Y \sim \cN
        \begin{pmatrix}
            \begin{pmatrix}
                Y \\ Y
            \end{pmatrix}
            ,
            \alpha \begin{pmatrix}
                \Sigma_{Y^*} & \Sigma_{Y^*, Y} \\
                \Sigma_{Y, Y^*} & \Sigma_{Y}
            \end{pmatrix}
        \end{pmatrix}
    \end{equation*}
    and use the estimate
    \begin{equation*}
        \widehat{\cov}^{*} = \tr\left((B-1)^{-1} \sum_{b=1}^B g(Y^{(b)})(Y^{*(b)} - \bar Y^{*(B)})^\top \right)
    \end{equation*}
    Then the analogous Efron and BY estimators would be
    \begin{equation*}
        \mathrm{Efr}_\alpha(g) = \|Y - g(Y)\|_2^2 + 2\widehat{\cov} - 2\widehat{\cov}^* + \tr \Sigma_{Y^*} - \tr\Sigma_Y
    \end{equation*}
    \begin{equation}\label{eq:spby}
        \mathrm{BY}_\alpha(g) = \|Y - g(Y)\|_2^2 + \frac{2}{\alpha}\widehat{\cov} - \frac{2}{\alpha}\widehat{\cov}^* + \tr \Sigma_{Y^*} - \tr\Sigma_Y
    \end{equation}
    For similar reasons, it is not clear these will estimate the optimism at the noise-elevated level. We can see BY is successful for linear smoothers as 
    $$\cov(Y^*, g(Y)) = \tr(\Sigma_{Y ^*, Y} S^\top) = \alpha^{-1}\E\left[\widehat{\cov}^* \mid Y\right]$$
    However for more complicated models it seems harder to understand its behavior. Further, as noted by \citet{Oliveira2021}, BY can suffer higher variance when the prediction function $g$ is unstable, whereas estimators that smooth $g$ with added noise reduce this instability and therefore reduce estimator variance.
    \begin{figure} [ht]
        \centering
        \includegraphics[width=.95\linewidth]{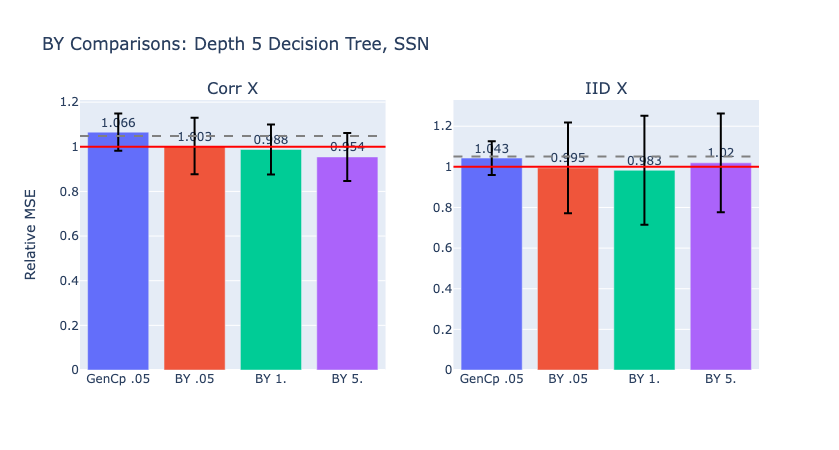}
        \caption{Relative MSE of depth 5 decision tree for the SSN model. `Corr X' refers to the usual way we gnerate $X$ described in \cref{sec:sims}. `IID X' refers to $X$ generated as IID draws from an isotropic Gaussian $\cN(0, I\sigma^2)$. `GenCp .05' refers to the estimator proposed in \cref{cor:corrgencp} with $\alpha = .05$. Labels of the form `BY u' refer to the BY estimator of \cref{eq:spby} with $\alpha = u$.}
        \label{fig:by_comp}
    \end{figure}
    \cref{fig:by_comp} illustrates the above described qualities of BY. We simulate as described in \cref{sec:sims}, only varying how we generate $X$ and $\beta$. We choose $p=s=200$. On the left is the usual construction detailed in \cref{sec:sims}, on the right we generate an uncorrelated $X$. We see that when $X$ is structured that BY underestimates both $\mathrm{Err}(g)$ and $\mathrm{Err}_\alpha(g)$ for larger $\alpha$. On the other hand when $X$ is IID, we see that although BY is unbiased for all values of $\alpha$, it also exhibits much higher variance for all values of $\alpha$.

\section{What is CV estimating under structured noise?} \label{sec:whatcv}
    With our approach to the problem, we can also provide some insight into what a particular CV method is estimating in spatial settings and thus when it may be suitable to employ.
    
    We begin with some simple insights from decomposing the MSE. Here we depart from our usual setting of NSN or SSN. Instead we investigate the more typical setting for CV methods, out-of-sample error. Examining an estimation fold $E$ and prediction fold $P$, we see that for a linear smoother $S$
    $$
    \E\|Y - S(X_E, X_P) Y\|_{\Theta_P}^2 = \E\|\mu_P - S(X_E, X_P) Y\|_{\Theta_P}^2 + \tr(\Sigma_{Y} \Theta_P) - 2\tr(S(X_E, X_P)\Sigma_Y \Theta_P)
    $$
    If we assume our model is (approximately) unbiased, we see that
    $$\E\|Y - S(X_E, X_P) Y\|_{\Theta_P}^2 = \tr(M_{P,E})$$
    for some $M_{P,E} \in \R^{n \times n}$ which is a function of only observable quantities and the (assumed known) covariance. The bias of a single fold from a CV method is due to its $\tr(M_{P,E})$ not matching that of the true training and test sets.Thus we can simulate this for single splits $E, P$ from various CV methods to get some insight into when their estimates match up with the error they aim to estimate.

\begin{figure}[H]
        \centering
        \begin{subfigure}[c]{.48\textwidth}
            \includegraphics[width=\textwidth, height=\textwidth]{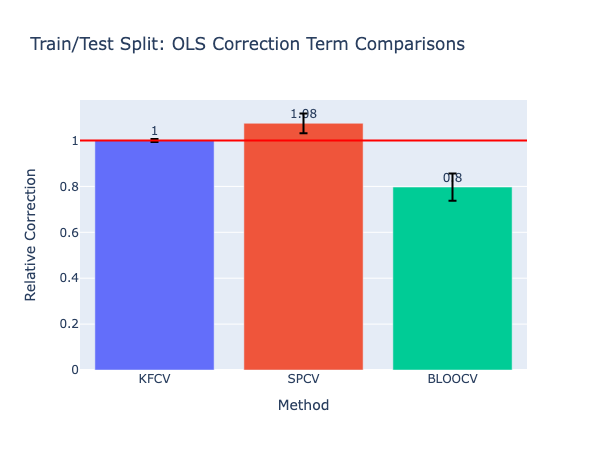}
        \end{subfigure} 
        \begin{subfigure}[c]{.48\textwidth}
            \includegraphics[width=\textwidth, height=\textwidth]{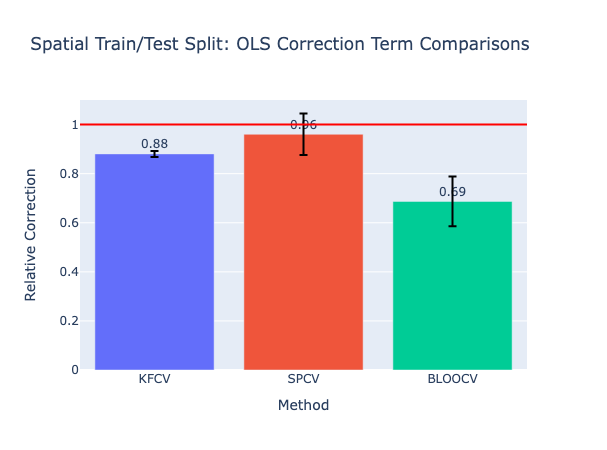}
        \end{subfigure}
        
        \caption{Relative correction term $M_{P,E}$ for various CV methods compared to different train/test splits. `KFCV' refers to one split of $k$-fold CV, and `SPCV' denotes one split of spatial CV with groups assigned by $k$-means, both with $k$=5. `BLOOCV' refers to one split of BLOOCV with buffer radius 10.}
        \label{fig:cv_comps}
    \end{figure}

    \cref{fig:cv_comps} illustrates two such comparisons in the case of OLS. We plot the correction terms of one fold of several CV methods relative to some train/test split of interest. Namely, we look at a standard random split as well as a spatial split of the data into train and test sets. We then apply KFCV, SPCV, and BLOOCV to the training set to get one CV estimation/prediction fold. The average correction term with standard error bars, normalized by the correction term of the true train/test split, is plotted. The data $(X,Y)$ are generated as described in \cref{sec:sims} with the following differences. We generate $n=2500$ samples placed in a grid on a $50\times 50$ space. We set $p=s=30$. For smoothing the spikes in columns of $X$, we use a Matern kernel with $\nu=.5, \rho=5.$
    These initial results point to CV methods providing good estimates of error when the folds in CV emulate the training and testing sets. We provide the following theorem to formalize this intuition in a more general setting. The result is stated for one split, but directly extends to CV. The proof is deferred to \cref{sec:cvproof}.
    \begin{theorem}\label{thm:cvest}
        Consider some $d$-dimensional study area of interest $R \subseteq \R^d$ that has features $X_r \in \R^p$ and response $Y_r \in \R$ associated with each point $r \in \R$. Suppose we have a mechanism of generating sample data $F_{\text{sample}} : \Omega \times \N \times \N \to 2^{R} \times 2^{R}$ to sample points from $R$, and a mechanism specifying how a user would sub-sample data $F_{\text{user}}: 2^{R} \times \N \times \N \to 2^{R} \times 2^{R}$. Consider sampled sets
        \begin{gather*}
            T_{tr}(\omega), T_{ts}(\omega) = F_{\text{sample}}(\omega, n_{tr}, n_{ts}), \quad 
            E(\omega), P(\omega) = F_{\text{user}}(T_{tr}(\omega), n_e, n_p), \quad
        \end{gather*}
        Assume that $ F_{\text{sample}}, F_{\text{user}}$ are such that, for some $\omega' \ed \omega$, $T_{tr}(\omega'), T_{ts}(\omega') = F_{\text{sample}}(\omega', n_e, n_p)$ we have $E(\omega) \ed T_{tr}(\omega')$ for all $n_e, n_p$. Further, define $r_{ts}$ as a simple random draw from $T_{ts}(\omega)$ and similarly let $r_p$ be a simple random draw from $P(\omega)$. Assume $r_{ts} \ed r_p$. Let $Y, r_{ts}, r_p$ be such that $Y_{r_{ts}}, Y_{r_p}$ have finite second moments. Consider some general model $g$ such that $g(X_{tr}, Y_{tr}, X_{r_{ts}}) \cp g_\infty(X_{r_{ts}})$ as $n_{tr} \to \infty$ for some $g_\infty$ such that $g(X_{E}, Y_{E}, X_{r_{p}}) \cp g_\infty(X_{r_{p}})$. Additionally let $g$ satisfy
        \begin{gather*}
            \E[g(X_{tr}, Y_{tr}, X_{r_{ts}})^2] \to \E[g_\infty(X_{r_{ts}})^2], \quad \E[g(X_{E}, Y_{E}, X_{r_{p}})^2] \to \E[g_\infty(X_{r_{p}})^2]
        \end{gather*}
        Then,
        $$\lim_{n_{tr} \to \infty} \E\left[n_{ts}^{-1}\|Y_{ts} - g(X_{tr}, Y_{tr}, X_{ts})\|_2^2 \right] = \lim_{n_e \to \infty} \E\left[n_p^{-1} \|Y_P - g(X_E, Y_E, X_P)\|_2^2 \right]$$
    \end{theorem}
    Intuitively this says that a practitioner should choose a CV method that splits the training set into folds such that the distribution of the folds matches the distribution of your training and test sets.

    This theorem can also help us explain the results presented in \citep{Wadoux2021}. They found regular $k$-fold CV to provide unbiased estimates in spatial settings when the training set comes from a simple random sample of locations. Since a simple random sample of a simple random sample is still a simple random sample, the unbiasedness of $k$-fold CV in this setting is unsurprising in light of the above theorem. However, when \citet{Wadoux2021} simulated the training set from a clustered sampling of locations, none of the CV methods unbiasedly estimated prediction error. This is obvious in light of the analysis and discussion in this section. When the original sample is a clustered sample, none of the CV methods subsample points in a way that replicates the distribution of the training set or test set locations, and thus don't estimate the error accurately.

\section{Simulations} \label{sec:sims}
    We now detail the setup for our simulation examples. Unless otherwise specified, simulation dataset $(X,y)$ is generated in the following way. For sample size $n=100$, features $p=200$, and sparsity $s=5$.
    We begin by generating $X$ and $\beta$ by the following procedure:
    \begin{enumerate}
        \item Place samples equally spaced on a $10 \times 10$ square with locations $(c_x, c_y) \in [0,10] \times [0,10]$.

        \item Independently generate columns $X_i$ by a two-step process:
        \begin{enumerate}
            \item Randomly select $2\log(n)$ locations and place spikes at those locations drawn from a $\mathrm{Unif}([-3,-1]\cup[1,3])$ distribution.
        
            \item Smooth spikes to interpolate values at all other locations using a Matern kernel. We use Euclidean distance and parameters $\nu=.5, \rho=1.$
        \end{enumerate}
        
        \item Draw $\beta_i \sim \mathrm{Unif}[-1,1]$ for $i=i_1,\dots,i_s$ for randomly picked indices $i_j$. The remaining elements of $\beta$ are set to 0.
        
    \end{enumerate}
    
    From this point forward, $X, \beta$ are held fixed. Then for each iteration of a simulation, we generate
    $$\begin{pmatrix}
        Y \\ Y^*
    \end{pmatrix} = \begin{pmatrix} X\beta \\ X\beta
    \end{pmatrix} + \epsilon$$
    for noise $\epsilon \sim \cN(0, \Sigma)$. $\Sigma$ will be generated differently depending on the setting: NSN or SSN.

    For NSN, that is where $Y \indep Y^* \mid X$, $\Sigma$ is block diagonal where both blocks are equal to $\Sigma_{Y} = \delta\Sigma_S + (1-\delta)I$ for $\delta=.75$. 
    \begin{itemize}
        \item $\delta$ controls the relative amounts of structured and measurement noise. Usually we choose $\delta = .75$.

        \item $\Sigma_S$ is generated from a Matern kernel using Euclidean distance and parameters $\nu=2.5, \rho=5.$
    \end{itemize}

    For SSN, that is where $Y \not\indep Y^* \mid X$, then $\Sigma$ has the same diagonal blocks, and the covariance between $Y^*,Y$ is simply $\delta\Sigma_S$.
    
    In either case, the covariance matrix is then scaled to achieve a signal-to-noise ratio (SNR) of $0.4$. That is we scale $\Sigma$ so that $\var_n(X\beta) / \Sigma_{ii} = 0.4$ where $\var_n$ denotes the empirical variance of the vector $X\beta$.

    In some simulations we employ train/test splits. What is referred to as a random split is simple random sampling of locations. What we refer to as spatial splits generated by a two-step clustered-sampling approach:
    \begin{enumerate}
        \item Split the space into a $5\times 5$ grid of sub-squares.

        \item For training proportion $p_{tr}$, sample $2p_{tr}$ fraction of sub-squares.

        \item Sample $n p_{tr}$ locations uniformly from the selected sub-squares to create the training set. All unsampled points become the test set.
    \end{enumerate}
    This creates clustered samples, replicating what is often found in spatial datasets.For such simulations, we use the corresponding $\Theta_e,\Theta_p$ constructions described in \cref{sec:mallows}. That is, our error metric is the MSE on the test locations only and for $g$ trained only on the training locations.

    Figures include:
    \begin{itemize}
        \item A red line for the ground truth target $\mathrm{Err}$ estimated by Monte-Carlo. We plot MSE relative to the ground-truth, so the red line will always be at 1.
        
        \item A dashed gray line for the noise-elevated target $\mathrm{Err}_\alpha$ at $\alpha=.05$, also estimated by Monte-Carlo.

        \item Error bars, where present, to show standard errors.
    \end{itemize} 
    Common labels for different estimation methods in figures are:
    \begin{itemize}
        \item `GenCp' refers to the variation of GC estimator appropriate for the model in question.

        \item `KFCV' refers to standard $k$-fold CV with $k=5$. 

        \item `SPCV' refers to spatial CV based on groups determined by $k$-means clustering on locations with $k=5$.

        \item `BY', where present, refers to the generalization of the Breiman-Ye estimator laid out in \cref{sec:optim}.

        \item `Split', where present, refers to using the train/test split (on $Y$) as an estimate of $\mathrm{Err}(g)$, which is test locations MSE on $Y^*$ for $g$ trained on the training locations of $Y$.
    \end{itemize}
    We use a variety of prediction models in our simulations. We detail their parameter settings here:
    \begin{itemize}
        \item The relaxed lasso is fit with $\lambda=.31$. This $\lambda$ was chosen because it is close to the middle of the solution path.

        \item Decision trees are fit to depth 3. 

        \item RF models are fit with 100 trees of depth 3.

        \item The lasso is fit with $\lambda=.31$. This $\lambda$ was chosen because it is close to the middle of the solution path.

        \item LassoCV is the lasso with with $\lambda$ tuned by CV. We tune on a grid of 10 values logarithmically spaced between .01 and 10.

        \item RidgeCV is the ridge with with $\lambda$ tuned by CV. We tune on a grid of 10 values logarithmically spaced between .01 and 10.
    \end{itemize}

    The overarching takeaway from these various simulations is that GC is always unbiased when the covariance is known, and appear to only suffer at most a small bias when the covariance is estimated ($\sim 5\%$).

    \subsection{Adaptive linear smoothers}

        \begin{figure}[H]
            \centering
            \includegraphics[width=.95\linewidth]{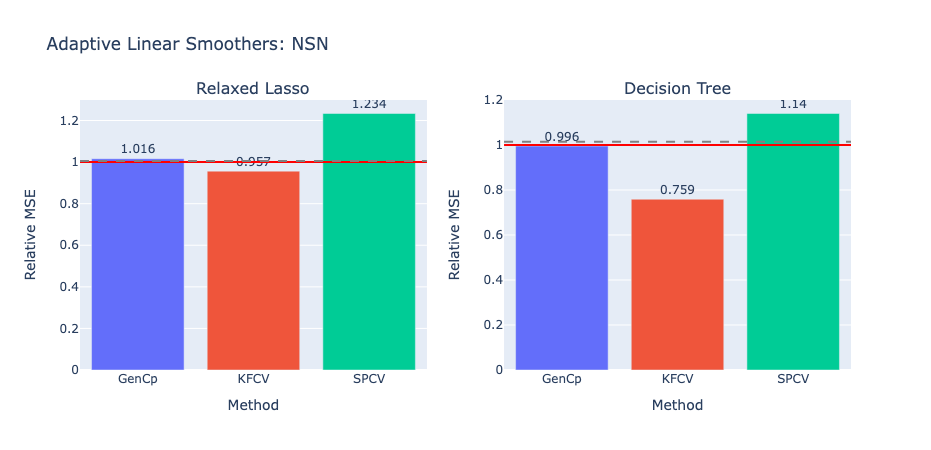}
            \caption{Relative MSE of adaptive linear smoother prediction functions for the NSN model. We see KFCV underestimates the MSE, while SPCV overestimates the MSE, while GenCp appears unbiased.}
            \label{fig:alin_ind}
        \end{figure}

        \begin{figure}[H]
            \centering
            \includegraphics[width=.95\linewidth]{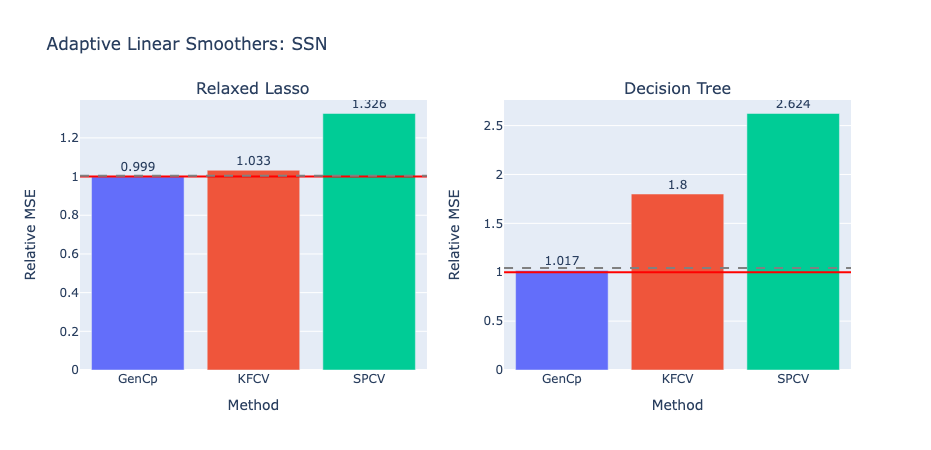}
            \caption{Relative MSE of adaptive linear smoother prediction functions for the SSN model. In this correlated case, we see both KFCV and SPCV overestimate MSE, while GenCp appears unbiased.}
            \label{fig:alin_corr}
        \end{figure}

    \subsection{Bagged adaptive linear smoothers}
        For these simulations we chose $n=400, p=s=5$ and the following modifications. We used a non-linear mean function, the Friedman function $\mu(x) = (10\sin(\pi x_1x_2) + 20(x_3 - 0.5)^2 + 10x_4 + 5x_5)/6$ from \citep{Friedman1991}. We also applied random and spatial 25/75 train/test splits.
        \begin{figure}[H]
            \centering
            \begin{subfigure}[c]{.48\textwidth}
                \includegraphics[width=\textwidth]{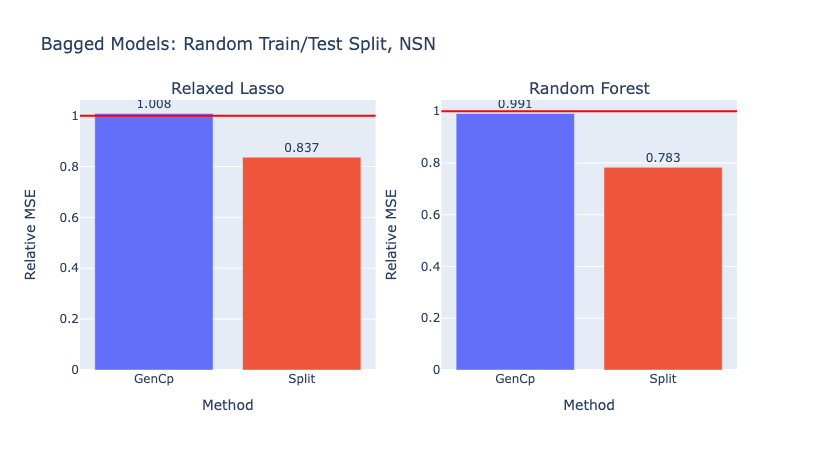}
            \end{subfigure} 
            \begin{subfigure}[c]{.48\textwidth}
                \includegraphics[width=\textwidth]{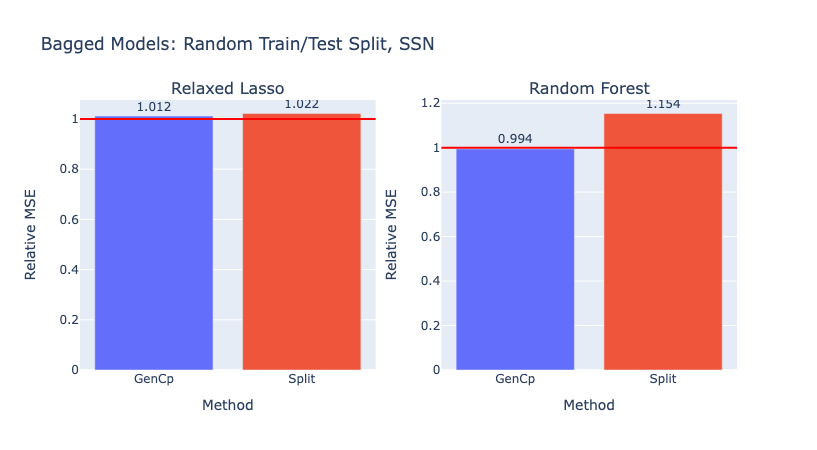}
            \end{subfigure}
            \begin{subfigure}[c]{.48\textwidth}
                \includegraphics[width=\textwidth]{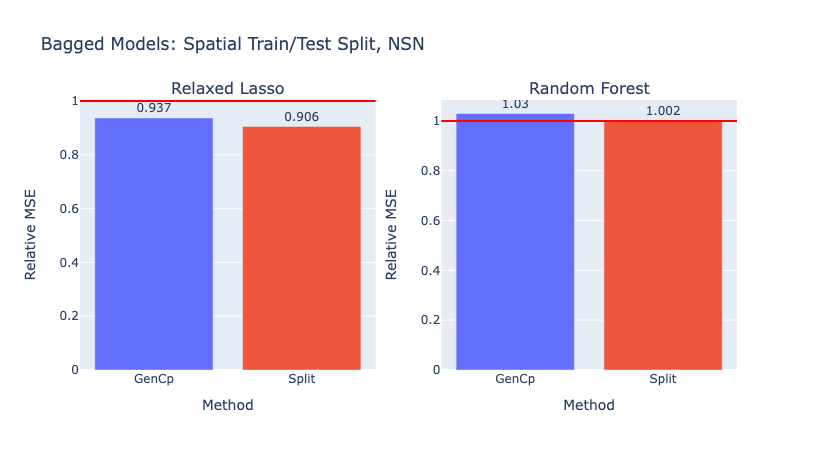}
            \end{subfigure} 
            \begin{subfigure}[c]{.48\textwidth}
                \includegraphics[width=\textwidth]{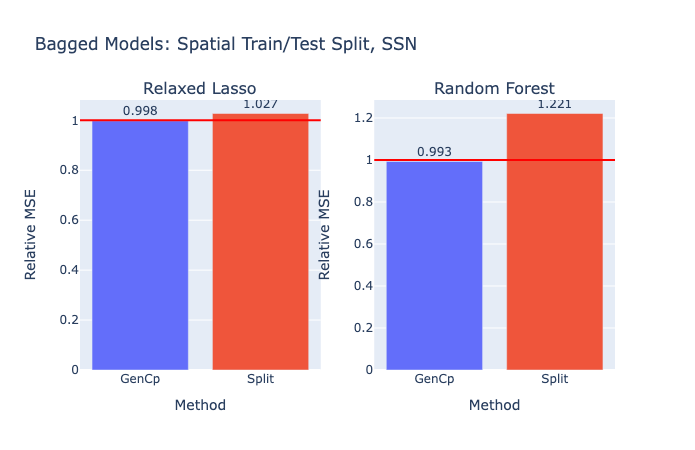}
            \end{subfigure}
            
            \caption{Relative MSE of bagged prediction functions. Here `Split' refers to the naive MSE from the train/test split on original response $Y$. The red line is the MSE on the new sample $Y^*$.}
            \label{fig:bag_sims}
        \end{figure}

    \subsection{General models}

    \subsection{Estimating covariance}
        We set $n=400, p=s=5$. We used a non-linear mean function $\mu$. In particular, we used the Friedman function $\mu(x) = (10\sin(\pi x_1x_2) + 20(x_3 - 0.5)^2 + 10x_4 + 5x_5)/6$ \citep{Friedman1991}. We see that although we do lose estimation accuracy when we estimate the covariance, it is limited to about 5\%.

    \subsection{Random vs trace correction}
        We set $n=400, p=200, s=5$. 
        \begin{figure}[H]
            \centering
            \includegraphics[width=.95\linewidth]{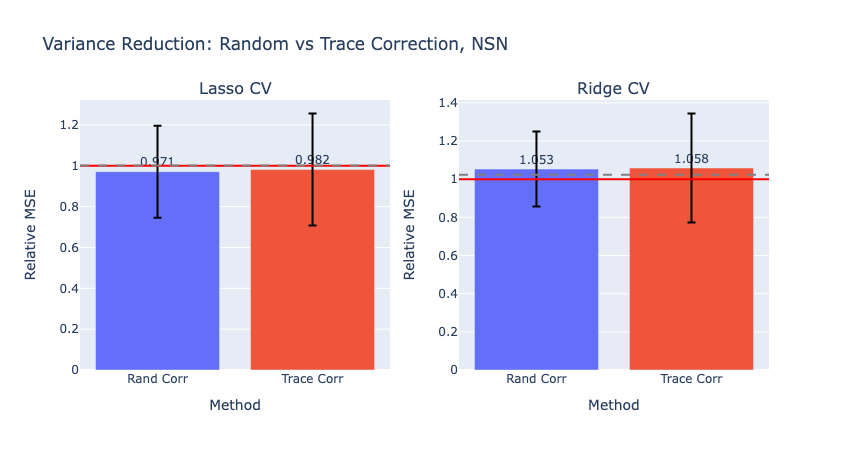}
            \caption{Comparing variance of using random correction vs trace correction for lasso and ridge tuned by cross-validation. }
            \label{fig:rand_vs_trace}
        \end{figure}
        \cref{fig:rand_vs_trace} shows that the random correction has smaller standard error bars in both cases while maintaining unbiasedness. Therefore it is more desirable than the trace correction.
    
\section{Application: Nevada geothermal data} \label{sec:realdata}
    In this application, we make use of data from one of our industry collaborators, Zanskar Geothermal and Minerals. They provided us with geological and temperature data for the state of Nevada in the USA. In particular, the dataset has 1,009 sampled locations with temperature data, and each location has 21 geological features measured for it as well. Examples of features are the depth at which the measurement was taken, gravity density, magnetic density, geodetic strain rate, quaternary fault distance. We elide the details of these features, trusting that our geology colleagues have measured features predictive of the geological temperature.

    We have two goals. The first is to replicate, on real data, simulation results showing CV methods do disagree with GC, suggesting the inefficacy of CV methods in spatial settings. The second goal is to demonstrate how one could use GC estimates as a criteria for model selection.

    We fit a RF model using 100 trees of depth 2. We then produce error estimates using Gen $C_p$, KFCV, and SPCV with standard error bars under the SSN model. Given our theoretical and simulation results, this suggests CV methods overestimate MSE on this Nevada dataset.
    \begin{figure}[H]
        \centering
        \includegraphics[width=.75\linewidth]{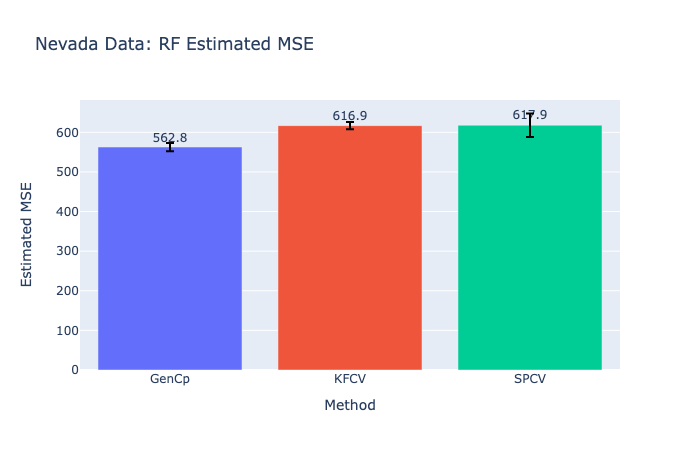}
        \caption{Comparing estimated MSE of various estimation methods. We see that KFCV and SPCV estimate the error to be significantly higher than our GenCp method.}
        \label{fig:nv_ests_comp}
    \end{figure}

    For model comparison, we compare various models and parameter settings:
    \begin{itemize}
        \item `XGB2': An XGBoost model with 100 estimators of depth 2.
        
        \item `XGB6': An XGBoost model with 100 estimators of depth 6.
        
        \item `RF2': An RF model with 100 estimators of depth 2.
        
        \item `RF6': An RF model with 100 estimators of depth 6.

        \item `ENetCV': An elastic net model tuned by CV.
    \end{itemize}
    For each model, we compute in-sample MSE on the full Nevada dataset.
    \begin{figure}[H]
        \centering
        \includegraphics[width=.75\linewidth]{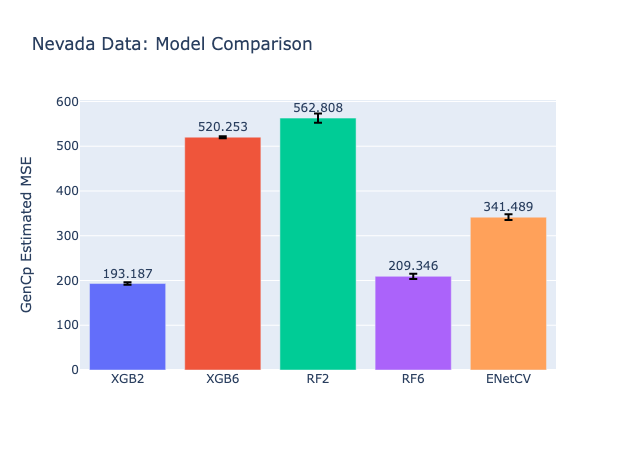}
        \caption{Comparing estimated in-sample MSE of various prediction models.}
        \label{fig:nv_model_sel}
    \end{figure}
    \cref{fig:nv_model_sel} shows the results on the Nevada dataset. We see that XGB2 and RF2 performed the best and second best, respectively. Thus GC suggests XGB2 would be the best predictor of new temperature measurements at these locations.
    
\section{Conclusion} \label{sec:conclusion}
    We have proposed a method for unbiasedly estimating the error of arbitrary predictive models on spatially correlated Gaussian data. We also present an approach to bag any predictive model and provide unbiased estimates of its prediction error. This approach is appealing because it allows one to bootstrap in a way that respects the covariance structure of the data, and is computationally less expensive than applying \cref{cor:corrgencp} with $g$ being the bagged model. 

    After this analysis, we use our setup to provide some insight into an ongoing debate around the use of various CV methods for prediction error estimation. We see that train/test error estimates depend on the covariance of both the training samples with the predictions as well as the testing samples with the predictions. This suggests how to use and interpret CV estimates is heavily dependent on the sampling procedure and ensuring that aligns with the CV procedure used. For example, if future data are going to be sampled at random from the study area, then traditional k-fold CV better emulates this sampling process and should provide better estimates than a spatial approach.

    \subsection{Predicting local maxima}
        In many settings, the practitioner may not want to know the error on average over new locations drawn from some distribution, but instead want to know the model performance at a local maxima of its prediction. Now the choice of prediction locations depends on the training locations via the model predictions. GC treats the training and test locations as fixed \textit{a priori}, and thus not immediately applicable. Devising a CV splitting strategy that would replicate this procedure, then perhaps \cref{thm:cvest} would apply, but such a strategy does not seem immediately obvious.
        
    \subsection{Incorporating into CV}
        
        Essentially our method allows one to properly do train/test splits for spatial Gaussian data. One could imagine incorporating this into a CV scheme, where you use this method to compute multiple estimates of prediction error on different folds and averaging them. What exactly this CV scheme would be estimating is not immediately clear and would depend on the CV scheme used. However, if done such that each train/test split is identically distributed, then as in the IID case, it would offer a better estimate of predictive error than a single split. A more thorough investigation of such an approach would be interesting. 

    \subsection{Estimating covariance}
        As noted in \cref{sec:estcov}, we find that using the same prediction function for covariance estimation and prediction error estimation works well in our simulations. However, it would be interesting to investigate this phenomena and derive some theory justifying this choice.

    \subsection{Other distributions}
        Our focus has been correlated Gaussian data, however it would be impactful to be able to derive analogous results for other distributions and models, such as common exponential families and their corresponding canonical generalized linear models (GLMs). \citet{Oliveira2021} note that you need only two independent views of your data $Y$. In \citet{Oliveira2022} they work out the IID Poisson case. In most cases we cannot fission our data as we can with Gaussians or Poissons, so a different approach may be necessary.

    \subsection*{Acknowledgments}
    KF is thankful to James Yang and Chen Cheng for useful discussions and feedback. KF was supported by the National Science Foundation Graduate Research Fellowship under Grant No. DGE-1656518.

\bibliographystyle{apalike}
\bibliography{refs}
\newpage
\appendix

\section{Proofs of unbiasedness of estimators}
    \subsection{Proof of \texorpdfstring{\cref{thm:alincp}}{NSN adaptive linear smoother GC}} \label{sec:alinproof}
        The proof is a simple algebra exercise leveraging the independence of $W^\perp$ and $W$
        \begin{align*}
            &\ \E\left[\|Y^* - SY \|_{\Theta_p}^2 \mid S(W)=S, X\right] - \E \left[ \|W^\perp - SY \|_{\Theta_p}^2 \mid S(W)=S, X \right] \\
            =&\ \tr(\Theta_p\Sigma_Y) - \tr(\Theta_p \Sigma_{W^\perp}) - 2\E \left[ (W^\perp - \mu)^\top \Theta_p(\mu - SY) \mid S(W)=S, X \right] \\ 
            =&\ \tr(\Theta_p\Sigma_Y) - \tr(\Theta_p \Sigma_{W^\perp}) + 2\E \left[ (W^\perp - \mu)^\top \Theta_p SY \mid S(W)=S, X \right] \\ 
            =&\ \tr(\Theta_p\Sigma_Y) - \tr(\Theta_p \Sigma_{W^\perp}) + 2\E \left[ (W^\perp - \mu)^\top \Theta_p SA^\perp(W^\perp-\mu) \mid S(W)=S, X \right] \\ 
            =&\ \tr(\Theta_p\Sigma_Y) - \tr(\Theta_p \Sigma_{W^\perp}) + 2\tr(\Theta_p SA^\perp\Sigma_{W^\perp}) \\
            =&\ \tr(\Theta_p\Sigma_Y) - \tr(\Theta_p \Sigma_{W^\perp}) + 2\tr(\Theta_p S\Sigma_{Y})
        \end{align*}
        The proof is analogous for $S(W)W$ but in that case the last term is 0 since $W^\perp \indep W$ by construction.

    \subsection{Proof of \texorpdfstring{\cref{thm:alincorrcp}}{SSN adaptive linear smoother GC}}\label{sec:alincorrproof}
        Noting that $\E[N^\perp] = \E[N]$, we have
        \begin{align*}
            &\ \E\left[\|Y^* - SY \|_{\Theta_p}^2 \mid S(W) = S, X\right] - \E\left[\|N^\perp - \Delta \|_{\Theta_p}^2 \mid S(W) = S, X \right] \\
            =&\ \E\left[\|N - \Delta \|_{\Theta_p}^2 \mid S(W) = S, X \right] - \E\left[\|N^\perp - \Delta \|_{\Theta_p}^2 \mid S(W) = S, X \right] \\
            =&\ \tr(\Theta_p (\Sigma_N - \Sigma_{N^\perp})) - 2\E\left[(N^\perp - \E[N])^\top \Theta_p (\E[N] - \Delta) \mid S(W) = S, X \right] \\
            =&\ \tr(\Theta_p (\Sigma_N - \Sigma_{N^\perp})) + 2\E\left[(N^\perp - \E[N])^\top \Theta_p (S-\Gamma)Y \mid S(W) = S, X  \right] \\
            =&\ \tr(\Theta_p (\Sigma_N - \Sigma_{N^\perp})) + 2\E\left[(W^\perp - \mu)^\top (I - \Gamma)^\top \Theta_p (S-\Gamma)A^\perp (W^\perp - \mu) \mid S(W) = S, X  \right] \\
            =&\ \tr(\Theta_p (\Sigma_N - \Sigma_{N^\perp})) + 2\tr\left((I - \Gamma)^\top \Theta_p (S-\Gamma)A^\perp \Sigma_{W^\perp} \right) \\
            =&\ \tr(\Theta_p (\Sigma_N - \Sigma_{N^\perp})) + 2\tr\left((I - \Gamma)^\top \Theta_p (S-\Gamma) \Sigma_Y \right)
        \end{align*}
        Then noting the first trace term on the rhs can be replaced with $$\tr(\Sigma_N - \Sigma_{(I - \Gamma)Y}) - \|(I - \Gamma)\omega\|_{\Theta_p}^2 / \alpha$$
        gives the desired result.
        
        Analogous steps give the result for refitting with $W$, the only difference being the last term is 0.

    \subsection{Proof of \texorpdfstring{\cref{thm:bagging}}{NSN bagged models GC}} \label{sec:baggingproof}
        Let $Z_K$ denote the matrix of residuals from our $K$ estimators, and $\bar Z_K$ be the vector of average residuals, averaged across estimators. There is a dependence here on the choice of refit vector, but it should be obvious from context and we suppress that in our notation. In the case of refitting with $Y$, these quantities would be
        $$
        \mathbb{R}^{n \times K} \ni Z_K = 
        \begin{pmatrix} \vert & \vert & \vert \\
        \cdots & Y^* - S(W_k)Y & \cdots \\ \vert & \vert & \vert \end{pmatrix},
        \qquad \R^n \ni \bar Z_K = Y^* - \frac{1}{K} \sum_{k=1}^K S(W_k)Y
        $$
        Now we can state our result. In matrix notation we can write
        $$
        \left\| Y^* - \frac{1}{K} \sum_{i=1}^K S(W_k)Y \right\|_{\Theta_p}^2 = \tr(\bar Z_K^\top \Theta_p \bar Z_K)
        $$
        It is a simple algebra exercise in expanding out $\tr\left( (Z_K^{\perp})^\top \Theta_p Z_K^\perp\right)$ to see that
        \begin{equation} \label{eq:bagdecomp}
            K\tr(\bar Z_K^\top \Theta_p \bar Z_K) = \tr(Z_K^\top \Theta_p Z_K) - \tr\left( (Z_K^{\perp})^\top \Theta_p Z_K^\perp\right)
        \end{equation}
         Inspecting the right hand side of \cref{eq:bagdecomp}, we see the first term is the sum of prediction errors for each individual estimator, and thus estimable by \cref{thm:alincorrcp}. The second term is just a function of the centered predictions, which are observable. Taking expectations on both sides gives the result.
         
         To see the analogous result for refitting with $W_k$, simply replace $S(W_k)Y$ with $S(W)W_k$ in the definitions of $Z_K$, $\bar Z_K$, and $Z_K^\perp$.

\section{Parametrics v. nonparametric random forest} \label{sec:rfcomps}
    Here we present an additional simulation comparing the MSE of the usual nonparametric bootstrapped RF and the parametric bootstrapped RF. We set $n=400, p=s=30$ with locations placed in a grid on a $20 \times 20$ square.
    \begin{figure}[H]
        \centering
        \includegraphics[width=.95\linewidth]{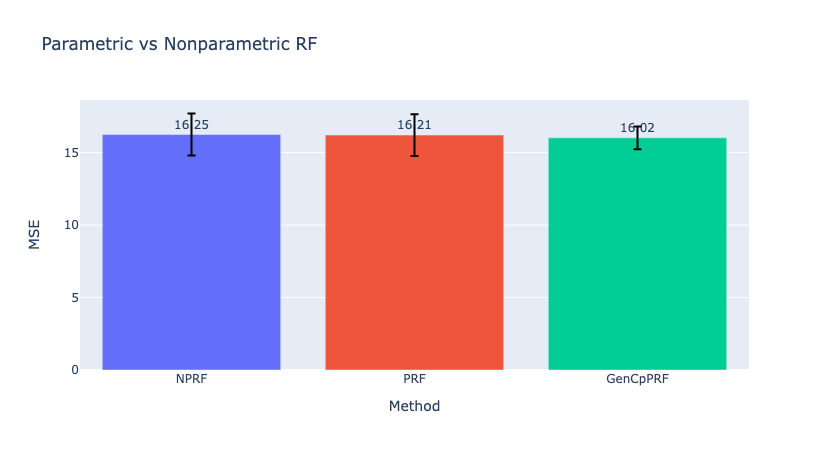}
        \caption{Comparing estimated MSE of parametric vs nonparametric random forests (RF). `NPRF' is ground-truth error of the nonparametrically bootstrapped RF, that is the standard RF. `PRF' is ground-truth error of the parametrically bootstrapped RF. `GenCpPRF' is the estimated error of the parametrically bootstrapped RF using the estimator proposed in \cref{cor:corrgencp}.}
        \label{fig:p_vs_np_rf}
    \end{figure}
    \cref{fig:p_vs_np_rf} shows that the parametric bootstrap RF has very similar MSE to the usual nonparametric bootstrap RF. Verifying our theory, we see our generalized $C_p$ estimator for parametric bagged models also agrees with both. Thus we have some confidence in using the parametric bagged error estimates for the nonparametric RF to achieve significant computational savings compared to the fully general version of the estimator.

\section{Proof of \texorpdfstring{\cref{prop:smooth}}{smoothness of noise-elevated target}} \label{sec:smoothproof}
    This proof is almost a one-for-one replication of the proof given in \cite{Oliveira2021} in the IID case, merely changing the quadratic form. However we provide the proof for the non-IID case here for completeness. The proposition is a direct consequence of the following lemma taking $f(w) = \|\mu - g(w)\|_{\Theta_p}^2$ and noting that $\E \|Y^* - g(W)\|_{\Theta_p}^2 = \E\|Y^* - \mu\|_{\Theta_p}^2 + \E \|\mu - g(W)\|_{\Theta_p}^2$ since $Y^* \indep W$.
    \begin{lemma}
        For $\alpha \ge 0$ let $W_\alpha \sim \cN(\mu, \Sigma_{W_\alpha})$ where
        $$\Sigma_{W_\alpha} = (1 + \alpha) \Sigma_Y$$
        Let $f: \R^n \to \R$ such that for some $\beta > 0$ and integer $k \ge 0$,
        $$\E\left[f(W_\beta) \|W_\beta - \mu\|_{\Sigma_{Y}^{-1}}^{2m} \right] < \infty, \qquad m=0,\dots, k$$
        then $\alpha \mapsto \E[f(W_\alpha)]$ has $k$ continuous derivatives on $[0,\beta)$.
        \begin{proof}
            We begin by proving continuity. Fix $\alpha \in [0,\beta)$. Note that
            \begin{align*}
                \lim_{t \to \alpha} \E[f(W_t)] &= \lim_{t\to\alpha} \int \frac{f(w)}{(2\pi(1+t))^{n/2} |\Sigma_{Y}|^{1/2}} \exp\left\{ -\frac{1}{2(1+t)}\|w - \mu\|_{\Sigma_{Y}^{-1}}^2 \right\} dw\\
                &= \int \lim_{t\to\alpha} \frac{f(w)}{(2\pi(1+t))^{n/2} |\Sigma_{Y}|^{1/2}} \exp\left\{ -\frac{1}{2(1+t)}\|w - \mu\|_{\Sigma_{Y}^{-1}}^2 \right\} dw\\
                &= \E[f(W_\alpha)]
            \end{align*}
            where in the second line we use dominated convergence theorem (DCT). This is justified as
            \begin{align*}
                \frac{f(w)}{(2\pi(1+t))^{n/2} |\Sigma_{Y}|^{1/2}} \exp\left\{ -\frac{1}{2(1+t)}\|w - \mu\|_{\Sigma_{Y}^{-1}}^2 \right\} &\le \frac{f(w)}{(2\pi)^{n/2} |\Sigma_{Y}|^{1/2}} \exp\left\{ -\frac{1}{2(1+\alpha)}\|w - \mu\|_{\Sigma_{Y}^{-1}}^2 \right\} \\
                &\le \frac{f(w)}{(2\pi)^{n/2} |\Sigma_{Y}|^{1/2}} \exp\left\{ -\frac{1}{2(1+\beta)}\|w - \mu\|_{\Sigma_{Y}^{-1}}^2 \right\} \\
            \end{align*}
            with the last line being integrable by assumption. The proof the first derivative, which generalizes to any higher derivative, relies on the Leibniz integral rule for the derivative
            $$\frac{\partial}{\partial \alpha} \E[f(W_\alpha)] = -\frac{n}{2(1+\alpha)}\E[f(W_\alpha)] + \frac{1}{2(1+\alpha)^2} \E\left[f(W_\alpha) \|W_\alpha - \mu\|_{\Sigma_Y^{-1}}^2\right]$$
            Then the integrands of these expectations are bounded by 
            $$\frac{f(w)}{(2\pi)^{n/2} |\Sigma_{Y}|^{1/2}} \|w - \mu\|_{\Sigma_Y^{-1}}^{2m} \exp\left\{ -\frac{1}{2(1+\beta)}\|w - \mu\|_{\Sigma_{Y}^{-1}}^2 \right\}$$
            for $m=0,1$ respectively. Thus by DCT we have continuity of the derivative on $[0,\beta)$. This general approach applies for higher derivatives using the Leibniz integral rule and DCT.
        \end{proof}
    \end{lemma}

\section{Proof of \texorpdfstring{\cref{thm:cvest}}{CV theorem}} \label{sec:cvproof}
    This is a simple result from swapping limits using Vitali's convergence theorem. Recall that Vitali's convergence theorem states that if $f_n \cp f$ and $\E[f_n^2] \to \E[f^2]$, then $f_n \overset{\cL^2}{\to} f$. For $f_n = Y_{r_{ts}} - g(X_{tr}, Y_{tr}, X_{r_{ts}})$ and $f = Y_{r_{ts}} - g_\infty (X_{r_{ts}})$, they satisfy the conditions of the theorem by our assumptions on $g$, thus we can conclude
    $$\lim_{n_{tr} \to \infty} \E[\|Y_{r_{ts}} - g(X_{tr}, Y_{tr}, X_{r_{ts}})\|_2^2] = \E[\|Y_{r_{ts}} - g_\infty (X_{r_{ts}})\|_2^2]$$
    $$\lim_{n_{e} \to \infty} \E[\|Y_{r_{p}} - g(X_{E}, Y_{E}, X_{r_{p}})\|_2^2] = \E[\|Y_{r_{p}} - g_\infty (X_{r_{p}})\|_2^2]$$
    Thus we have
    \begin{align*}
        \lim_{n_{tr} \to \infty} \E\left[n_{ts}^{-1}\|Y_{ts} - g(X_{tr}, Y_{tr}, X_{ts})\|_2^2 \right] &= \lim_{n_{tr} \to \infty} \E\left[\|Y_{r_{ts}} - g(X_{tr}, Y_{tr}, X_{r_{ts}})\|_2^2 \right] \\ 
        &= \E\left[\|Y_{r_{ts}} - g_\infty(X_{r_{ts}})\|_2^2 \right]\\ 
        &= \E\left[\|Y_{r_{p}} - g_\infty(X_{r_p})\|_2^2 \right]\\ 
        &= \lim_{n_e \to \infty} \E\left[n_p^{-1} \|Y_P - g(X_E, Y_E, X_P)\|_2^2 \right]
    \end{align*}
    where the second to last equality follows from the fact that $r_{ts} \ed r_p$ by assumption, and thus $Y_{r_{ts}} - g_\infty(X_{r_{ts}}) \ed Y_{r_{p}} - g_\infty(X_{r_{p}})$.

\end{document}